\documentclass[showpacs,twocolumn,aps,preprintnumbers,letterpaper]{revtex4}
\usepackage{amsmath,amssymb}
\usepackage{epsfig}
\usepackage{graphicx}
\usepackage{amsmath}
\usepackage{slashed}
\usepackage{amsfonts}
\usepackage{epstopdf}
\usepackage{color}
\usepackage[caption=false]{subfig}
\usepackage[pdftex, pdfborder= 0 0 0, citecolor=blue, urlcolor=blue, linkcolor=blue, colorlinks=true, bookmarksopen=true]{hyperref}

\newcommand{\be}{\begin{equation}}
\newcommand{\ee}{\end{equation}}
\newcommand{\bear}{\begin{eqnarray}}
\newcommand{\eear}{\end{eqnarray}}
\newcommand{\ba}{\begin{array}}
\newcommand{\ea}{\end{array}}

\def\be{\begin{eqnarray}}
\def\ee{\end{eqnarray}}
\def\bea{\be}
\def\eea{\ee}

\def\vk{{\vec k}}

\def\roughly#1{\mathrel{\raise.3ex\hbox{$#1$\kern-.75em%
\lower1ex\hbox{$\sim$}}}}

\definecolor{davecolor}{rgb}{0.95,  0.5,  0.2}

\def\({\left(}
\def\){\right)}
\def\[{\left[}
\def\]{\right]}
\def\<{\langle}
\def\>{\rangle}

\def\tr{\mathop{\rm tr}}

\newcommand\half{{\ensuremath{\frac{1}{2}}}}
\newcommand\p{\ensuremath{\partial}}

\newcommand{\bwt}{\begin{widetext}}
\newcommand{\ewt}{\end{widetext}}

\newcommand{\bi}{\begin{itemize}}
\newcommand{\ei}{\end{itemize}}
\newcommand{\ben}{\begin{enumerate}}
\newcommand{\een}{\end{enumerate}}
\newcommand{\bca}{\begin{cases}}
\newcommand{\eca}{\end{cases}}
\newcommand{\bln}{\begin{align}}
\newcommand{\eln}{\end{align}}
\newcommand{\bst}{\begin{split}}
\newcommand{\est}{\end{split}}

\newcommand\al{{\alpha}}

\newcommand\sig{\sigma}

\newcommand\om{\omega}

\newcommand\ga{{\ensuremath{{\gamma}}}}
\newcommand\Ga{{\ensuremath{{\Gamma}}}}

\newcommand\da{{\dagger}}

\newcommand\ov{\over}
\newcommand\ha{{\half}}

\def\le{\left}
\def\ri{\right}

\newcommand\sD{{\ensuremath{{\mathcal D}}}}

\newcommand\bpsi{{\bar \psi}}

\newcommand\ut{{\underline{t}}}
\newcommand\ur{{\underline{r}}}
\newcommand\ui{{\underline{i}}}

\newcommand\psinorm{\boldsymbol{\psi}}

\newcommand\Phinorm{\boldsymbol{\Phi}}

  \long\def\comment#1{ }

  \newcommand{\beq}{\begin{eqnarray}}
  \newcommand{\eeq}{\end{eqnarray}}
  
 \def\simge{\mathrel{%
   \rlap{\raise 0.511ex \hbox{$>$}}{\lower 0.511ex \hbox{$\sim$}}}}
\def\simle{\mathrel{
   \rlap{\raise 0.511ex \hbox{$<$}}{\lower 0.511ex \hbox{$\sim$}}}}

\begin{document}

\title{Deep Inelastic Scattering on an Extremal RN-AdS Black Hole II:\\
Holographic Fermi Surface}

\author{Kiminad A. Mamo}
\email{kiminad.mamo@stonybrook.edu}
\affiliation{Department of Physics and Astronomy, Stony Brook University, Stony Brook, New York 11794-3800, USA}

\author{Ismail Zahed}
\email{ismail.zahed@stonybrook.edu}
\affiliation{Department of Physics and Astronomy, Stony Brook University, Stony Brook, New York 11794-3800, USA}



\date{\today}
\begin{abstract}
We consider deep inelastic scattering (DIS) on a dense nucleus described as an
 extremal RN-AdS black hole with holographic quantum fermions in the bulk. We evaluate the 1-loop fermion contribution to the R-current
on the charged black hole, and map it on scattering off a Fermi surface  of a dense and  large nucleus  with fixed atomic number.
Near the black hole horizon, the geometry is that of AdS$_2\times $R$^3$ where the fermions develop an emergent  Fermi surface
with anomalous dimensions. DIS scattering off these fermions yields to anomalous  partonic distributions mostly at large-x, as well as modified hard scattering rules. The pertinent R-ratio for the black hole is discussed. For comparison, the structure functions and the R-ratio in the probe or dilute limit with no back-reaction on the geometry, are also derived. We formulate  a hybrid holographic model for DIS scattering on  heavy and light nuclei,  which compares favorably to the existing data for Pb, Au, Fe, C and He
over  a wide range of parton-x.
\end{abstract}


\maketitle

\setcounter{footnote}{0}


\section{Introduction}

Many years ago the EMC collaboration at CERN has revealed that DIS scattering on an iron nucleus deviates
substantially from deuterium~\cite{E665} contrary to established lore. Since then, many other collaborations
using both electron and muon probes  have confirmed this observation~\cite{NMC,SLAC,BCDMS}. Although
the nucleus is a collection of loosely bound nucleons with confined quarks, DIS scattering is much richer in
a nucleus. The nuclear structure functions display shadowing at low-x, a depletion at intermediate-x,  and an
enhancement due mostly to  Fermi motion at large-x.

QCD supports the idea that hadrons are composed of quarks and gluons as revealed by DIS scattering
of electrons on nucleons at SLAC. The scaling laws initially reported follows from scattering on point-like
object or partons. Because of asymptotic freedom, the partons interact weakly at short
distances leading to relatively small scaling violations at intermediate-x.  At low-x, perturbative QCD
predicts a large enhancement in the nucleon structure functions due to the rapid growth of the gluons
~\cite{BK} that eventually saturate~\cite{GW}. This observation has been confirmed at HERA~\cite{H1,ZEUSS}.

DIS in holography at moderate-x is different from weak coupling as it involves hadronic  and
not partonic constituents~\cite{POL}. The large gauge coupling causes the charges to rapidly deplete
their energy and momentum, making them invisible to hard probes.
However, because the holographic limit enjoys approximate conformal
symmetry,  the form factors exhibit various scaling laws including the
parton-counting rules~\cite{BF}.  The holographic structure functions fail to reproduce
the Callan-Gross sum rule~\cite{POL} at intermediate-x, but agree with it at large-x when
the parton momentum fraction neighbors 1~\cite{Braga:2011wa}.
In contrast, DIS scattering at low-x on a non-extremal
thermal black-hole was argued  to be partonic  and fully saturated~\cite{HATTA}.

This paper is a follow up on our recent investigation of DIS scattering on a nucleus as an extremal
RN-AdS black hole~\cite{Mamo:2018ync}. In the double limit of a large number of colors and gauge coupling,
the leading contribution amounts to the Abelian part of the R-current being absorbed in bulk by the
black-hole. After mapping at the boundary, the ensuing nuclear structure functions show strong
shadowing at low-x, but wane exponentially for large-x as originally noted for the thermal black hole in~\cite{HATTA}.

At next to leading-order, the R-current scatters off charged
fermionic pairs forming a holographic Fermi liquid around the black hole. The purpose of this paper
is to detail DIS scattering on this  dense holographic liquid as the analogue of DIS scattering on a nucleus
described as a Fermi liquid. Some aspects of this liquid near the horizon were initially discussed in lower dimensions~\cite{MANY}.
It should be noted that at next to leading order the R-current scatters also off bulk charged scalars. However,
these scalars are bosonic and do not form a Fermi surface. Their contribution will not be considered in this work.

Standard DIS scattering  on a nucleus is mostly on
a Fermi gas in a mean field $^{\prime\prime}$trap$^{\prime\prime}$, so the present calculations show how the same scattering operates on an emergent
Fermi surface with strongly coupled fermionic constituents in a  $^{\prime\prime}$trap$^{\prime\prime}$ produced by a charged black-hole.
A chief observation is that the partonic structure functions at large-x, are modified  by an
emergent Fermi surface. The latter follows from an AdS$_2\times$R$^3$ reduction of the geometry near the  black-hole horizon,
and asymptotes a warped Fermi liquid near the boundary.
The  corresponding R-ratio exhibits  shadowing at very low-x, anti-shadowing at
intermediate-x and Fermi motion at large-x, much like  the R-ratio for DIS scattering on finite nuclei. Shadowing is caused by
the coherent many-body effects and is captured by DIS scattering in leading order on  the black-hole at low-x, while Fermi motion at large-x is due to the
incoherent scattering on  quantum fermions around the  black hole  in the form of a  holographic Fermi surface.

This paper consists of several new results:
1/ an explicit derivation of the structure functions for DIS scattering on the emerging holographic Fermi surface
around an extremal black hole;  2/ the characterization of these structure functions both at large-x and low-x,
with the identification of new anomalous exponents at large x;
3/ an explicit derivation of the R-ratio for DIS scattering on the extremal black hole as a model for DIS scattering
on a dense and finite nucleus; 4/ an explicit derivation of the same structure functions in the probe fermion limit as a model
for DIS scattering on a dilute nucleus; 5/ a comparative study of the R-ratio in the probe limit.; 6/ a detailed comparison to
the empirical DIS scattering data from light to heavy nuclei.

The organization of the paper is as follows: in section II, we briefly review the setting for the
extremal RN-AdS black hole, and the key characteristics of the holographic Fermi liquid.
In section III,  we derive the contribution to the boundary  effective action of an R-photon
scattering off bulk quantum fermions. The result is quantum and dominant at large-x, and 
corrects the classical and leading contribution from the bulk black hole.
In section IV, we analyze the contribution stemming from the quantum
fermions near the horizon.  In section V, we
detail our derivation of the R-ratio for DIS scattering on a dense nucleus as quantum corrected holographic
black hole.  For comparison,
we discuss in section VI the probe or dilute limit with the bulk fermions carrying a finite density
in AdS without affecting the underlying geometry. The pertinent R-ratio in this regime is
derived and analyzed. In section VII we motivate a hybrid holographic model for DIS scattering
on light and heavy nuclei which compares favorably to the existing world-data for a wide range
of parton-x.  Our conclusions are in section VIII. Some useful details are found in
several  Appendices.



\section{Extremal black hole : dense limit}

\begin{figure}[!htb]
 \includegraphics[height=10cm]{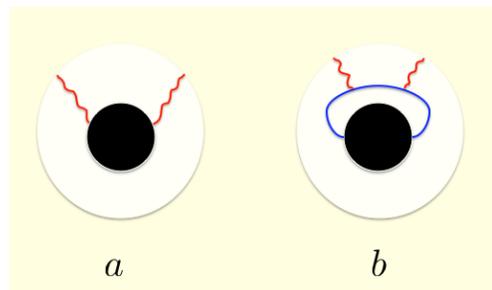}
  \caption{Absorptive part of the R-current  on a nucleus as an extremal RN-AdS black hole: (a) absorptive  contribution
  to order $N_c^2$; (b) absorptive fermionic contribution to order $N_c^0$.}
  \label{scattering}
\end{figure}

In this work we will address DIS scattering on a cold and dense nucleus as a dual to an RN-AdS black hole following on our
recent analysis~\cite{Mamo:2018eoy}. Conventional DIS scattering on cold nuclei
with many of the conventions used are reviewed in~\cite{REVIEW}.  In holography,
DIS scattering on a nucleus as an RN-AdS black hole is illustrated in Fig.~\ref{scattering}. In the holographic limit, the leading contribution is Fig.~\ref{scattering}a with the structure functions being the absorbed parts of the R-current. To this order, the
structure functions have considerable support mostly at low-x~\cite{Mamo:2018ync} (see below).  At next-to-leading order, the R-current is absorbed through the
virtual fermionic loop shown in Fig.~\ref{scattering}b. This loop describes a fermionic hallow around the RN-AdS black hole
that acts as a holographic Fermi liquid. Below we detail how this contribution leads to structure functions with mostly support at large-x.
This description is complementary to our recent analysis based on a generic density expansion around a trapped Fermi liquid~\cite{Mamo:2018eoy}.

\subsection{The extremal and charged  black hole}

The RN-AdS black hole is described by effective gravity coupled to a U(1) gauge field in a
5-dimensional curved AdS space~\cite{adsBH}

  \be
  S=\frac 1{2\kappa^2}\int d^5x\sqrt{-g}\,(\mathbb R-2\Lambda) -\frac 1{4e^2}\int d^5x\sqrt{-g}F^2
  \label{RNADS}\nonumber\\
  \ee
 The Ricci scalar is $\mathbb R$, and $\kappa^2=8\pi G_5$ and $\Lambda=-6/R^2$ are the gravitational and cosmological
  constant.  The curvature radius of the AdS space is $R$ with line element

 \be
  ds^2=\frac {r^2}{R^2}\left(-f\, dt^2+d\vec{x}^2\right)+\frac {R^2}{r^2 f}dr^2
  \label{MADS}
  \ee
and warping factor

\be
 \label{FR}
 f(r)=\left(1-\frac{r_{+}^2}{r^2}\right)\left(1-\frac{r_{-}^2}{r^2}\right)\left(1+\frac{r_{+}^2}{r^2}+\frac{r_{-}^2}{r^2}\right)
 \ee
with $r_+>r_-$ the outer-inner horizons satisfying $f(r_\pm)=0$.


The black-hole is charged and sources the R-potential

  \be
  A_t=\mu-\frac Q{r^2}
  \label{GA}
  \ee
provided that the electric charge $Q$ and the geometrical charge $q$ satisfy
\be
  \frac{q^2R^2}{Q^2}=\frac 43 \times\frac{2\kappa^2}{4e^2}=\frac{R^2}{6\tilde{\alpha}}\,,
  \label{CHARGE}
  \ee
where
\begin{eqnarray}
   2\kappa^2=&&\frac{8\pi^2 R^3}{N_c^2}\nonumber\\
   4e^2=&&\tilde{\alpha}\frac{64\pi^2 R}{N_c^2}
   \label{EOS}
   \end{eqnarray}
We have defined  $\tilde{\alpha}=1$  for a U(1) R-charge,  and
$\tilde{\alpha}=\frac{1}{4}\frac{N_c}{N_f}$ for a D3-D7  U(1) vector charge.
The temperature of the RN-AdS black hole is

  \be
  T=\frac{r_+^2f'(r_+)}{4\pi R^2}=\frac{r_+}{\pi R^2}\Big(1-\frac{\mu^2\pi^2 R^4\gamma^2}{r_+^2}\Big)
  \label{TEMP}
  \ee
with  $\gamma^2={1}/{12\pi^2\tilde{\alpha}}$. The chemical potential $\mu$ is fixed by the zero potential condition
on the outer horizon $A_t(r_+)=0$ or $\mu=Q/r_+^2$. At extremality where $T=0$, we have $r_+=r_-=\pi R^2\gamma\mu=\frac{R^2}{2\sqrt{3}}\sqrt{\tilde{\alpha}}$.

\subsection{Holographic Fermi liquid}

The fermionic fields in bulk are  characterized by the Dirac action in a charged AdS black hole geometry

\be
\label{Dirac}
 S = - \int d^{5} x \sqrt{-g} \, i (\bar \psi \Ga^M \sD_M \psi  - m \bpsi \psi)
 \ee
with $\bpsi = \psi^\da \Ga^\ut$, and the long derivative

 \be
 \sD_M  = \p_M + {1 \ov 4} \om_{ab M} \Ga^{ab} - i e_R A_M
 \ee
 The indices $M , N \cdots$ or $\mu , \nu, r \cdots$  refer to the space-time indices, and $a,b, \cdots$ to
 space-time indices with underline correspond to tangent space indices. Therefore, for example, $\Ga^a$ denotes the gamma matrices in the tangent space, $\Ga^M$ denotes gamma matrices in the curved spacetime.  They are specifically given in Appendix A.

A bulk fermion field of mass $m$  and  R-charge $e_R$ is dual to a composite boundary field of
conformal dimension $\Delta=\frac 32 +mR$. Since the horizon of the extremal charged RN-AdS
black is characterized by a finite U(1) electric field, fermionic pair creation takes place through the Schwinger
mechanism. As a result, the black-hole say with positive
R-charge absorbs the negative part of the pairs and expel the positive part. Since AdS is hyperbolic
and confining, the positive charge falls back to the surface of the black hole, accumulating into a hallow
 or holographic Fermi liquid.

The characteristics of the low-lying excitations of the holographic Fermi  liquid for low frequencies
$|k^0|<\mu$ and low momenta $k=|\vec k|$, have been discussed
in~\cite{Faulkner:2009wj, Gubser:2012yb}. In particular,
near the horizon the AdS$_5$ geometry factors
into AdS$_2\times$ R$^3$.
\\
\\
{\bf Case-1: $e_R^2\tilde{\alpha}<\frac{1}{4}(mR)^2$}
\\
\\
The fermions exhibit  strong distorsion
in the AdS$_2$ geometry, with~\cite{Faulkner:2009wj}
\be
\label{GR11}
{\cal G}_R^{11}(k^0, \vec k)=C(\vec k)\,(k^0)^{2\nu_k}\begin{pmatrix}
0 & 0\\
0 & 1
\end{pmatrix}\,,
\ee
and
\be
\label{k2kR}
\nu_k=&&\frac{\sqrt{\tilde{\alpha}}}{\mu}\left(k^2-k_R^2\right)^{\frac 12}\nonumber\\
\equiv &&
\frac{\sqrt{\tilde{\alpha}}}{\mu}\left(k^2-\frac{\mu^2}{3\tilde{\alpha}}\left(e_R^2\tilde{\alpha}-\frac{1}{4}(mR)^2\right)\right)^{\frac 12}\,.
\ee
Note that $k_R^2<0$ in this case, i.e., for $e_R^2\tilde{\alpha}<\frac{1}{2}(mR)^2$.

Throughout, we will use the block notation to refer to the fermionic retarted (Feynman) propagators

\be
{\cal G}_{R,F}=\begin{pmatrix}
{\cal G}^{11} & {\cal G}^{12}\\
{\cal G}^{21}& {\cal G}^{22}
\end{pmatrix}_{R,F}\,
\ee
\\
\\
{\bf Case-2: $e_R^2\tilde{\alpha}>\frac{1}{4}(mR)^2$}
\\
\\
 For  $k_R^2>0$ and  $k\leq k_R$, the corresponding
holographic spectral function exhibits oscillating behavior  and gapless excitations,
with comparable real and imaginary parts. In other words, the excitations in this oscillating region
are short lived as they form and quickly fall into the extremal RN-AdS black hole.

\begin{figure}[!htb]
 \includegraphics[height=10cm]{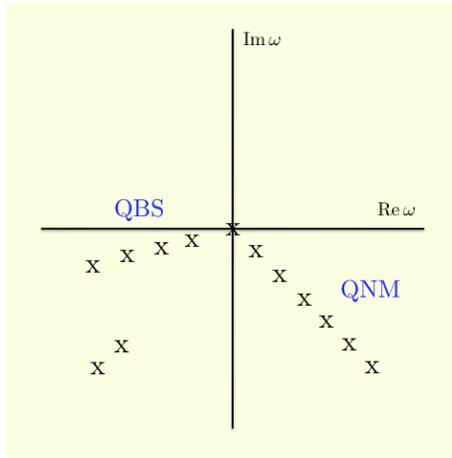}
  \caption{Schematic description of the poles of the Green function.
 For sufficiently large U(1) charge $e_R$ some of the quasi-normal modes  (QNM) of the RN-AdS
  black-hole transmute to  narrow quasi-bound states (QBS) by moving closer to the real-axis.}
  \label{fig_qbs}
\end{figure}

Further arguments~\cite{Faulkner:2009wj, Gubser:2012yb} show that the
fermionic density diverges near the horizon causing  strong back reaction. As a result,
the near horizon geometry becomes a Lifshitz geometry whereby the Fermi-like volume
is resolved into concentric Fermi spheres each describing heavy Fermions with narrow
widths, thereby explaining the gapless like excitations. This resolution
occurs only for $|k^0|/\mu\sim e^{-N_c^2}$ and resorbs for $|k^0|/\mu\sim N_c^0$.
\\
\\
{\bf Case-3: $e_R^2\tilde{\alpha}>\frac{1}{4}(mR)^2$}
\\
\\
 For $k_R^2>0$ and  $k\geq k_R$, localized and long lived fermionic states emerge  that are characterized by
a Fermi momentum $k_F>k_R$.  In this case, the
retarded propagator near the Fermi surface reads~\cite{Faulkner:2009wj, Gubser:2012yb}

\be
 \mathcal{G}^{11}_{R}(k^0,\vk) \approx  {h_1 \ov  k-k_{F} - \frac 1{v_F}k^0-\Pi(k^0)}\begin{pmatrix}
0 & 0\\
0 & 1
\end{pmatrix}
  \label{spinorG}
 \ee
with
\be
\label{phaseE}
\Pi(k^0)=&&h_2\, e^{i \ga_{k_F}}\, (k^0)^{2 \nu_{k}}\nonumber\\
\nu_k=&&\frac{\sqrt{\tilde{\alpha}}}{\mu}\left(k^2-k_R^2\right)^{\frac 12}\nonumber\\
\ga_k = && \arg\left( \Gamma(-2\nu_k) \le(e^{-2 \pi i \nu_k} - e^{- \frac{2 \pi}{\sqrt{3}} e_R \sqrt{\tilde{\alpha}}} \ri)\right)\,.\nonumber\\
 \ee
The coefficients $h_{1}\sim\frac{r_{-}}{R^2}$ and $h_{2}\sim \Big(\frac{r_{-}}{R^2}\Big)^{1-2\nu_{k_{F}}}$ can be computed numerically.
For $\nu_k>\frac 12$, $\nu_k=\frac 12$ or $\nu_k>\frac 12$ we have  a Fermi liquid, a marginal Fermi liquid
or  a non-Fermi liquid,  respectively. Note that the transition from a non-Fermi liquid to a Fermi liquid occurs for
$k^0\approx k^0_c$ which is fixed by the condition $k^0_c\approx |v_F\Pi(k^0_c)|$. Below, we will refer to the non-Fermi liquid
region $k^0<k^0_c$ as the near horizon or emergent contribution, and the Fermi liquid region $k^0>k^0_c$ as the far horizon or dilute contribution.
The two contributions will be added approximately since an exact construction that interpolates between these two limiting regimes is likely
numerical and outside the scope of this work.

 A schematic description
of the poles of (\ref{spinorG}) is given in Fig.~\ref{fig_qbs}. For sufficiently large effective charge  $e_R\sqrt{\tilde{\alpha}}$,  some of the largely damped quasi-normal
modes (QNM) of the RN-AdS black hole transmute to narrow quasi-bound states (QBS) close to the real axis for fixed
$k<k_F$.  For increasing $k\rightarrow k_F$ the narrow QBS start crossing the origin $\omega=0$ turning to equally
spaced  holographic  Fermi surfaces
(here 4 Fermi surfaces) as discussed in~\cite{Gubser:2012yb}.

For fermions with larger effective charge, i.e., for $e_R^2\tilde{\alpha}>\frac{1}{4}(mR)^2$ or $k_R^2>0$, pair creation takes place near the horizon as we noted earlier. A hallow of charged fermions
at the Fermi surface with $k_F>k_R>0$,
 that supports quasi-particles with ${\cal G}_F^{11}$ given in (\ref{spinorG}).
For hard R-probes with large $q^0$ in the DIS kinematics, only ${\cal G}_R^{11}(k^0, \vec k)$ is modified close to the horizon, since ${\cal G}_R^{11}(\omega_1, k+q)$ carries a large momentum and is mostly unmodified in the ultraviolet,

\begin{widetext}
\be
\label{A7}
{\rm Im}\,\mathcal{G}_R^{11}(\om_1,k+q)\, {\rm Im}\,\mathcal{G}_{R}^{11}(k^0,\vk)\rightarrow
{\rm Tr}\,\Big((\sigma_{1}(k^0+q^0)-i\sigma_2(k_x+q_x)-\omega_1)\pi\,\delta((k+q)^2+\omega_1^2)
\,{\rm Im}\,\mathcal{G}^{11}_{R}(k^0,\vk)\Big)
\ee
\end{widetext}

\section{Holographic Structure Functions}

The holographic structure functions on an extremal black-hole in leading order have been discussed in~\cite{Mamo:2018eoy},
to which we refer for further details.  For completeness, the results will be summarized below, and extended to allow for the
next to leading order contributions from the holographic Fermi liquid at the horizon.

\subsection{Structure functions}

We recall that the scattering amplitude of an R-photon of longitudinal momentum
$q^\mu=(\omega,q,0,0)$ scattering on a black-hole at rest in the Lab frame
with $n^\mu=(1,0,0,0)$, (\ref{D8}),  can be tensorially decomposed into  two invariant functions $\tilde{G}_{1,2}$~\cite{Mamo:2018ync}

\begin{widetext}
\be
\label{A1}
\tilde{G}^F_{\mu\nu}(q)=\left(\eta_{\mu\nu}-\frac{q_\mu q_\nu}{Q^2} \right)\tilde{G}_1+\left[n_\mu n_\nu-\frac{n\cdot q}{Q^2}
 (n_\mu q_\nu + n_\nu q_\mu)+
 \frac{q_\mu q_\nu}{(Q^2)^2}(n\cdot q)^2 \right] \tilde{G}_2\,\nonumber\\
 \ee
\end{widetext}
 with $Q^2=q^2$, thanks to with the current conservation and covariance.
The corresponding DIS structure functions for an R-photon on a black hole are defined as

\beq
\label{A2}
   \tilde{F}_1&\,=\,&\frac{1}{2\pi}\ {\rm Im} \tilde{G}_1, \nonumber\\[0.2cm]
  \tilde{F}_2&\,=\,&-\frac{(n\cdot q)}{E_{A}}\, {\rm Im} \frac 1{2\pi} \tilde{G}_2 \label{F12}\,.
  \eeq

 As in~\cite{Mamo:2018ync}, the rest frame of a cold and extremal  black hole will be dual to the rest frame of a cold nucleus
 at the boundary with fixed energy $E_A=\frac 34 A\mu$.
Since the binding energy in a nucleus is small, we also have $E_A\simeq Am_N$ and therefore
the chemical potential $\mu\simeq \frac 43 m_N$.
In our mapping, $m_N$ and $\mu$ are  interchangeble for estimates.  A hard photon
with virtual momentum $q^\mu= (\omega, q,0,0)$ scattering off the nucleus in the DIS kinematics satisfies
$q^2-\omega^2\equiv Q^2\rightarrow\infty$ with $\omega\simeq q$ and  fixed Bjorken-x

\be
\label{XA}
x_A=\frac{q^2}{-2q\cdot (nE_A)}\equiv \frac{Q^2}{2E_A\omega}=\frac{xm_N}{E_A}
\ee

\subsection{Classical black-hole in leading order: small-x}

As we noted earlier,
the leading order contribution to the structure functions (\ref{A2}) in DIS scattering is classical and of order $N_c^2$ as
illustrated in Fig.~\ref{scattering}. It does not involve scattering off the fermions near the holographic surface,
which is of order $N_c^0$. In the regime $Q\ll q\ll Q^2/\mu$ the leading contribution to the structure functions
vanishes,  as the probe spin-1 R-field is prevented from falling to the black-hole by an induced potential
barrier~\cite{HATTA}. The R-current correlator is purely real with an exponentially vanishing imaginary part.
In the regime $q\gg Q^3/\mu^2$, the barrier wanes away with the classical and leading  contribution
to the un-normalized structure function $\tilde F_2$ of the form~\cite{Mamo:2018ync}

\beq
  \label{FTL}
  &&\tilde F_2(x_A,Q^2)\approx \nonumber\\
  &&\tilde C_{T}\,\frac{\mu^2}{x_A}\,\left(\frac{x_A^2Q^2}{\mu E_A} \right)^{\frac 23}
  +\tilde C_{L}\frac{E_A}{\mu}\,\frac{\mu^2}{x_A}\,\left(\frac{x_A^2Q^2}{\mu E_A} \right)\nonumber\\
 \eeq
 with

 \be
 &&\tilde C_T=\frac{N_c^2}{2^{17/3}\pi^2\Gamma^2(1/3)\tilde{\alpha}^{5/3}}\nonumber\\
 &&\tilde C_L=\frac {N_c^2}{1152\pi^4\tilde \alpha^2}
 \ee
with $x_AE_A=xm_N$ and $C_L\ll C_T$. This result was shown to hold for
low-x or  $x_A\ll \sqrt{\mu E_A}/Q$, with the Callan-Gross relation-like $\tilde F_2=2x_A\tilde F_1$.
The normalized structure functions follow as~\cite{Mamo:2018ync}

\be
\label{F2BHX}
F_{1,2} \equiv 2E_AV_A\tilde{F}_{1,2}\equiv \left(\frac{12\pi\tilde\alpha A}{N_c\mu}\right)^2\tilde{F}_{1,2}
\ee
after using the black-hole equation of state. More specifically, we have
($Q^2=q^2>0$)

\be
\label{F2BH}
\frac{F_2^{\rm BH}(x,q^2)}A\approx \frac {C_T}x \left(\frac{3x^2q^2}{4m_N^2}\right)^{\frac 23}+
\frac{3C_L}{4x}\left(\frac{3x^2q^2}{4m_N^2}\right)\nonumber\\
\ee
with  $C_{T,L}/\tilde{C}_{T,L}=\pi^5(48\tilde\alpha)^2/2N_c^2$.

The normalization in (\ref{F2BHX})
amounts overall to  normalizing $\tilde F_{1,2}$ by the density of the black-hole,
canceling part of the model dependence of the equation of state. In a way, the
normalized $F_{1,2}$ are the  un-normalized black-hole
structure functions $\tilde F_{1,2}$ per degree of freedom.
(\ref{F2BH}) is dominated by the first
contribution at low-x. We now show
that the next contribution is dominated by scattering off bulk fermions at large-x
 from a holographic Fermi liquid close to the horizon.

\subsection{Classical black-hole in leading order: large-x}

The large-x contribution to the dense black-hole can be obtained using the WKB approach also developed for the thermal
black hole in~\cite{HATTA} with similar results modulo pertinent changes in the parameters and normalization. In particular,
the structure function for $x> \mu/q$ is  found to drop exponentially with the result

\be
\label{F2BHLX}
\frac{F_2^{\rm BH}(x,q^2)}A\approx
\frac{3C_L}{2x}\left(\frac{3x^2q^2}{4m_N^2}\right)D(x)\,,\nonumber\\
\ee
where to exponential accuracy

\be
D(x)\approx C_{D}e^{-\sqrt{\frac{q}{m_N}}\frac{m_N}{\mu}\sqrt{x}}\,,
\ee
with  $C_D\approx e^{-\frac{2\Gamma^2(1/4)\sqrt{6\tilde{\alpha}}}{3\sqrt{2\pi}}}$.
Below we will use (\ref{F2BH}) plus (\ref{F2BHLX}) to describe DIS scattering on the BH in leading order. We now proceed to analyze the
quantum and sub-leading correction.

\subsection{Quantum fermions in sub-leading order}

 The contribution of the sub-leading fermions to the
 induced effective action can be obtained through the holographic dictionary.
 It will be divided into two contributionds: 1/ the one stemming from the
 emergent Fermi surface near the horizon through the geometrical reduction
 to AdS$_2\times$R$^3$; 2/ the one stemming from its ultraviolet completion
 which is dual to a Fermi liquid in bulk AdS$_5$ which we will seek in the
 dilute approximation below.

 With this in mind, the shift of the R-field
 $A_M \to  A_0 \delta_{M0} + a_M$ amounts to a shift in the  Dirac action density in (\ref{Dirac}) at the
 origin of the minimal coupling of the R-field

 \be
 \label{D1}
 - i \overline \psi \le( - i e_Ra_\mu(r, q) \Ga^{\mu} \ri) \psi \equiv  \overline\psi B(r; q)\psi\nonumber\\
 \ee
In terms of (\ref{Dirac}-\ref{D1}) the bulk effective action for the 1-loop contribution in
Fig.~\ref{scattering}b at {\it zero temperature} reads

\begin{widetext}
 \be
 \label{D2}
{\cal S}_{F}[a_\mu] =-(-i)\int \frac{d^{4} q}{(2 \pi)^{4}}\frac{d^{4} k}{(2 \pi)^{4}}\,
 \int dr_1 \sqrt{g(r_1)} dr_2 \sqrt{g(r_2)}\, {\rm Tr}\Big(D_F(r_1,r_2;k+q) B(r_2; q) D_F(r_2,r_1; k)
B(r_1; q) \Big)
 \ee
\end{widetext}
The routing of the momenta in (\ref{D2}) corresponds to the hard fermion with $k+q$ and the
soft fermion with $k$.

The R-field in bulk $a(r,q)$ relates to the R-field at the boundary $A^{(0)}_\mu(q)$ through
the bulk-to-boundary propagator $K_{A}(r;q)$, which satisfies $K_{A}(r\rightarrow\infty;q)=1$,

\be
 \label{D4}
a_\mu (r; q) = K_A (r;q) A_\mu^{(0)} (q)\,.
 \ee
This allows the re-writing of (\ref{D2}) in the form of the boundary action

\begin{widetext}
 \be
 \label{D5}
{\cal S}_{F}[A^{(0)}_\mu] =&&\frac 12 \int \frac{d^{4} q}{(2 \pi)^{4}}A_\mu^{(0)}(q)A_\nu^{(0)}(-q)\nonumber\\
&&\times (-2i)\int \frac{d^{4} k}{(2 \pi)^{4}}\,
 \int dr_1 dr_2  \sqrt{g(r_1)} \sqrt{g(r_2)}\, {\rm Tr}\Big(D_F(r_1,r_2;k+q) Q^\mu(r_2; q) D_F(r_2,r_1; k)
Q^\nu (r_1; q) \Big) \nonumber\\
 \ee
\end{widetext}
with the dressed bulk vertices

\be
\label{D6}
 Q^\mu (r; q) &&=- i \big( - i  e_R \,K_A (r;q) \,\Gamma^{\mu}  \big) \nonumber\\
 && \approx -e_R \,\frac{qR^2}{r}\, K_1\Big(\frac{qR^2}{r}\Big)\,\Gamma^{\mu} \,.
 \ee
We have approximated the bulk-to-boundary $K_A(r;q)$ by its vacuum contribution, with
$K_1(x)$ the modified Bessel function.

In the DIS  regime $Q\ll q\ll Q^2/\mu$ with $Q^2=q^2$, the spin-$\frac 12$ fermion field remains localized
near the boundary as a potential barrier develops in bulk, a phenomenon also observed for spin-1 boson
fields~\cite{HATTA}. In this regime, we will approximate the hard part of the fermion propagator
by its  vacuum (in AdS$_5$) result~\cite{Gao:2009se}

\be
D_F(r_1,r_2;k+q) = \int d{\omega_1}\om_1 G_{F}(r_1,r_2;\omega_1,k+q)\,,\nonumber\\
\label{spectrep2}
\ee
where
\begin{equation}
  \label{specBB2}
G_{F}(r,r';\omega_1,k+q) = \psinorm_{\alpha}(r,\omega_1) \, \mathcal{G}_F^{\alpha \gamma} (\omega_1,k+q) \, \overline{\psinorm_{\gamma}} (r',\omega_1)
\end{equation}
with the vacuum (in AdS$_5$) solution~\cite{Polchinski:2002jw}

\be
\label{hwf1}
 && \psinorm_{1}(r,\omega_1)=\Big(\frac{R^2}{r}\Big)^\frac{5}{2}J_{mR-\frac{1}{2}}\Big(\omega_1\frac{R^2}{r}\Big)
 \begin{pmatrix}
0\\1
\end{pmatrix} \,\nonumber\\
  &&\psinorm_{2}(r,\omega_1)\equiv 0 \,,
\ee
 and

\begin{eqnarray}
\label{GF11F22}
 \mathcal{G}_F^{11} (\omega_1,k+q)=\frac{\sigma_{1}(k^0+q^0)-i\sigma_2(k_x+q_x)-\omega_1}{(k+q)^2+\omega_1^2-i\epsilon} \,, \nonumber\\
\mathcal{G}_F^{22} (\omega_1,k+q)=\frac{\sigma_{1}(k^0+q^0)+i\sigma_2(k_x+q_x)-\omega_1}{(k+q)^2+\omega_1^2-i\epsilon} \,. \nonumber\\
\end{eqnarray}

The soft part of the fermion propagator can be separated into its contribution deep in the infrared which is modified by the induced holographic
Fermi surface through the geometrical reduction to AdS$_2$$\times$R$^3$, and its ultraviolet completion as we noted earlier. More specifically, near the
AdS$_2$$\times$R$^3$ geometry, the infrared part of the soft the propagator is of the form

\be
\label{spinorG20}
D_{F}(r,r^\prime;k) =
\psinorm_{\al}(r,k^0,\vec{k})\mathcal{G}_{\al\beta}^{F}(k^0,\vec{k})\overline{\psinorm_{\beta}}(r',k^0,\vec{k})\nonumber\\
\ee
with
\be \label{spinorG2}
&& \mathcal{G}^{22}_{F}(k^0,\vk) = \nonumber\\
 &&\mathcal{G}^{11}_{F}(k^0,\vk\mapsto -\vk)=-\frac{1}{\mathcal{G}^{11}_{F}(k^0,\vk;m\mapsto -m)}\,.\nonumber\\
 \ee
Note that only $\mathcal{G}^{11}_{R}(k^0,\vk)$ has a singular or Fermi-like structure near $k\rightarrow k_{F}$.
Hence, we will ignore the contribution from $\mathcal{G}^{22}_{R}(k^0,\vk)$ to the current correlator. The normalizable
wave functions  are given in (\ref{swf}).


The time-ordered correlation function for the R-current follows from the functional derivative

\beq
\label{D7}
 \tilde{G}^{F\mu\nu}(q)\,=\,\frac{\delta^2 {\cal{S}_{F}}[A^{(0)}_\mu]}
 {\delta A^{(0)}_\mu(q)\delta A^{(0)}_\nu(-q)}\,,
 \eeq
Using the spectral  form of the Feynman propagator~(\ref{spinorG20}-\ref{spinorG2}), we can re-write (\ref{D7}) in a more compact form

 \begin{widetext}
\be
\label{D8}
\tilde{G}^{F{\mu\nu}}(q) = 2i
 \int \frac{d^{4} k}{(2 \pi)^{4}}
 \int d \om_1 \omega_1
{\rm Tr}\,\Big(\mathcal{G}^{\alpha\beta}_F (\om_1,k+q)\, \Lambda^\mu_{\beta\gamma} (\om_1;q;k) \,\mathcal{G}^{\gamma\delta}_{F}(k^0,\vk)  \,
 \Lambda^\nu_{\delta\alpha} (k;q;\om_1)\Big)\,,
  \nonumber\\                                                                                                                                                                                                                                                                                                                                                                                                                                                                                                                                                                                                                                                                                                                                                                                                                                                                                                                                                                                                                                                                                                                                                                                                                                                                                                                                                                                                                                                                                                                                                                                                                                                                                                                                                                                                                                                                                                                                                                                                                                                                                                                                                                                                                                            
  \ee
\end{widetext}
with the  dressed vertices

\bea \label{vertex11}
 &&\Lambda_{\beta \gamma}^\mu (\om_1;q;k)= \nonumber \\
 &&  \int dr_{2} \sqrt{g(r_{2})} \, \overline{\psinorm_\beta} (r_2,\om_1) \, Q^{\mu} (r_{2};q) \, \psinorm_\gamma (r_2,k)\,,\nonumber\\
\eea
and

\bea
 \label{vertex12}
 &&\Lambda_{\delta \alpha}^\nu (k;q;\om_1)= \nonumber \\
 &&  \int dr_{1} \sqrt{g(r_{1})} \, \overline{\psinorm_\beta} (r_{1},k) \, Q^{\nu} (r_{1};q) \,
 \psinorm_\alpha (r_1,\omega_1)\,.\nonumber\\
\eea

We recall that at zero temperature, the general Feynman and retarded propagators ${\cal G}_{F,R}$ are related by
the relationship

\be
\label{A4}
{\cal G}_F(k^0, \vec k)={\rm Re}\,{\cal G}_R(k^0, \vec k)+i\,{\rm sgn}(k^0)\,{\rm Im}\,{\cal G}_R(k^0, \vec k)\nonumber\\
\ee
Using (\ref{A4}) and the fact that ${\cal G}_F(k^0, \vec k)$ is analytic in the upper complex $k^0$-plane, allow for
the re-writing of the imaginary part of (\ref{D8}) in the form

 \begin{widetext}
 \be
  \label{GRx3}
{\rm Im}\,\tilde{G}^F_{\mu\nu}(q) &&=
\int \frac{d^4 k}{(2 \pi)^4}
\int d \om_1 \omega_1\,\Lambda_{\beta \gamma}^\mu (\om_1;q;k)\Lambda_{\delta \alpha}^\nu (k;q;\om_1)\,
{\rm Re}\,{ Tr}\,\Big({\mathcal{G}}_F^{\alpha \beta} (\om_1,k+q)\,\mathcal{G}^{\gamma \delta}_{F}(k^0,\vk)
 \Big)\,, \nonumber\\
 &&= \int \frac{d^3 k}{(2 \pi)^3}
 \int_{-|q^0|}^{0}\frac{dk^{0}}{2\pi}\int d \om_1 \omega_1\,\Lambda_{\beta \gamma}^\mu (\om_1;q;k)\, \Lambda_{\delta \alpha}^\nu (k;q;\om_1)
\,{\rm Tr}\,\Big({\rm Im}\,{\mathcal{G}}_R^{\alpha \beta} (\om_1,k+q)\,{\rm Im}\,\mathcal{G}^{\gamma \delta}_{R}(k^0,\vk)  \,
\Big)\,, \nonumber\\
\ee
\end{widetext}
This result shows that for $q^0=0$, the imaginary part vanishes as it should as the effective action
induced by the R-current (\ref{D5}) is real. For $q^0\neq 0$  this result  is clearly negative as it should,
since its contribution to (\ref{D5}) amounts to a self-energy for the R-field which amounts to damped
oscillations in time.

\subsection{Large-x near the horizon}

Using the vertex (\ref{vertex213}) for momenta near $k_F$, we can re-write (\ref{GRx3})

\be \label{GRx4}
&& {\rm Im}\,\tilde{G}^F_{xx}(q)=C^2(\nu_{k_{F}})C_{\theta}k_{F}^2\nonumber\\
&& \times(-1)\int_{0}^\infty\frac{d\om_1^2}{2} I_{z}^2(\om_1;q;k_F)\int_{k_R}^{k_F}dk\,a_+(k_0,k)^2\,{\rm Re}\, I_{k^0}(\om_1,q)\,,\nonumber\\
\ee
with

\begin{widetext}
\bea \label{I0}
{\rm Re}\, I_{k^0}(\om_1,q)&&={\rm Re}\, \int_{-\infty}^{\infty}\frac{dk^{0}}{2\pi}\,a_+(k_0,k)^2\,\tr\Big(\mathcal{G}_F^{11}(\om_1,k+q)\mathcal{G}^{11}_{F}(k^0,\vk)\Big)
\nonumber\\
&&={\rm Re}\,\int_{-|q^0|}^{0}\frac{dk^{0}}{2\pi}\,a_+(k_0,k)^2\,\tr\Big((\sigma_{1}(k^0+q^0)-i\sigma_2(k_x+q_x)-\omega_1)\pi\,\delta((k+q)^2+\omega_1^2)\,{\rm Im}\,\mathcal{G}^{11}_{R}(k^0,\vk)\Big)
\nonumber\\
&&\approx \,\int_{-|q^0|}^{0}\frac{dk^{0}}{2\pi}\,a_+(k_0,k)^2\,(-1)\omega_1\pi\,\delta((k+q)^2+\omega_1^2)\frac{\,{h_1 \rm Im}\,\Pi}{(k-k_{F}-\frac {k^0}{v_F}-{\rm Re}\,\Pi)^2+({\rm Im}\,\Pi)^2}
\eea
\end{widetext}
(\ref{I0}) can be simplified by enforcing the delta function

\begin{widetext}
\be \label{GRx5}
{\rm Im}\,\tilde{G}^F_{xx}(q)
&&\approx C^2(\nu_{k_{F}})C_{\theta}k_{F}^2\frac{\pi}{2}\int_{k_R}^{k_F}dk\,\int_{-|q^0|}^{0}\frac{dk^{0}}{2\pi}\,a^2_+(k_0,k)\,I_{z}^2(x_k;q;k_F)\,\sqrt{s_k}\frac{\,{h_1\rm Im}\,\Pi}{(k-k_{F}-\frac {k^0}{v_F}-{\rm Re}\,\Pi)^2+({\rm Im}\,\Pi)^2}\nonumber\\
&&\approx C_G(\nu_{k_F},z_{-})\,\Big(\frac{1}{q^2}\Big)^{\nu_{k_F}+\frac{3}{2}}\int_{k_R}^{k_F}dk\,k_{F}^2\,\int_{-|q^0|}^{0}\frac{dk^{0}}{2\pi}\nonumber\\
&&\times a^2_+(k_0,k)\,x_k^{\nu_{k_{F}}+5/2}(1-x_k)^{mR-1/2}\,_2F_1^2\Big(\frac{mR+\nu_{k_F}+2}{2},\frac{mR-\nu_{k_F}+1}{2},mR+\frac{1}{2},1-x_k\Big)\nonumber\\
&&\times\,\sqrt{s_k}\,\frac{\,{\rm Im}\,\Pi}{(k-k_{F}-\frac {k^0}{v_F}-{\rm Re}\,\Pi)^2+({\rm Im}\,\Pi)^2}\nonumber\\
\ee
\end{widetext}
with the overall constant

\be
&&C_G(\nu_{k_F},z_{-})=\frac{\pi}{2}\,z_{-}^{-(2\nu_k+2)}\,C^2(\nu_{k_{F}})\,C_{z}^2(\nu_{k_F})\,\tilde{h}_1(\nu_{k_{F}})C_{\theta}\,,\nonumber\\
\ee
where $\tilde{h}_1(\nu_{k_{F}})\equiv z_{-}h_{1}$ is a dimensionless constant to be determined numerically, and $z_{-}=\frac{R^2}{r_{-}}$.
Again, for the DIS kinematics we set
$x_k={-q^2}/{2k\cdot q}$,
and $|q^0|\approx q_x$.
We re-arranged the hypergeometric function $_2F_1$ using the same Pfaff identity (\ref{PFF}).
Note that for the special value of $\nu_{k}=mR+1$, one can see that the $x_k$ dependence of the integrand in (\ref{GRx5}) reduces to
the one in~\cite{Polchinski:2002jw} before the multiplication by the trace (for our case the trace is $\sqrt{s_k}$). However, for general $\nu_k$
the same partonic content as in (\ref{pdf}) is noted.

For narrow quasi-particles, we may use the substitution ($k_\perp=k-k_F$)

\be
\frac{\,{\rm Im}\,\Pi}{(k_\perp-\frac {k^0}{v_F}-{\rm Re}\,\Pi)^2+({\rm Im}\,\Pi)^2}\rightarrow
\pi \delta\left(k_\perp-\frac {k^0}{v_F}-{\rm Re}\,\Pi\right)\nonumber\\
\ee
and undo the $k^0$ integration in (\ref{GRx5})  with the result

\begin{widetext}
\be \label{GRxx5}
{\rm Im}\,\tilde{G}^F_{xx}(q)\approx \frac 12 \tilde{C}_G(\nu_{k_{F}})\int_{k_R}^{k_F}dk_\perp\,k_{F}^2\,\frac {a_+(k_0,k_{\perp})^2}{\left|\frac 1{v_F}+{\rm Re}\,\Pi^\prime\right|}
\left(\frac 1{q^2z_{-}^2}\right)^{\nu_{k_F}+1}
x_k^{\nu_{k_F}+2}(1-x_k)^{\tau-\frac 32}\,{}_2F_1^2\left(\tau_+, \tau_-, \tau-1, 1-x_k\right)\nonumber\\
\ee
\end{widetext}
with

\be
{\rm Re}\,\Pi^\prime=2h_2\nu_{k_F}|k^0|^{2\nu_{k_F}-1}\,{\rm Re}\left(e^{i\gamma_{k_F}}(-1)^{2\nu_{k_F}-1}\right)\nonumber\\
\ee
where  $k^0$ in $x_k$ is  solution to the transcendental equation

\be\label{transcendental}
k_\perp+\frac{|k^0|}{v_F}=h_2|k^0|^{2\nu_{k_F}}\,{\rm Re}\left(e^{i\gamma_{k_F}}(-1)^{2\nu_{k_F}}\right)\,,
\ee
and we have defined a dimensionless constant
\be
\tilde{C}_G(\nu_{k_{F}})&&= z_{-}^{2\nu_{k_{F}}+2}\times C_G(\nu_{k_{F}},z_{-})\nonumber\\
&&=\frac{\pi}{2}\,C^2(\nu_{k_{F}})\,C_{z}^2(\nu_{k_F})\,\tilde{h}_1(\nu_{k_{F}})\,C_{\theta}\,.\nonumber\\
\ee
Also note that $a_+(k_0,k_{\perp})=\overline{c}_1z_{-}\,k_{\perp}+\overline{c}_2z_{-}\,k_0$ is a dimensionless coefficient with dimensionless constants $\overline{c}_{1,2}=\tilde{c}_{1,2}/z_{-}$.

In arriving to (\ref{GRxx5}), we have made
use of  the Pfaff identity

\be
\label{PFF}
{}_2F_1(a,b,c,z)=(1-z)^{-a}\,_2F_1\left(a,c-b,c,\frac{z}{z-1}\right)\nonumber\\
\ee
In the holographic Fermi liquid,
the partonic distribution function is seen to develop a modified exponent. A  comparison of the partonic distribution
function (\ref{GRxwt7}) in the probe limit, to (\ref{GRxpure2}) where the black hole is present, translates

\begin{figure}[!htb]
 \includegraphics[height=6cm]{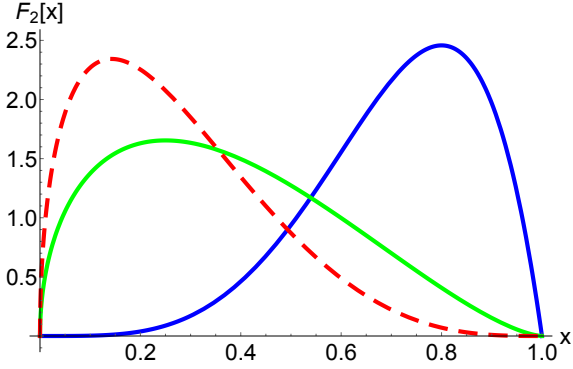}
  \caption{Large-x dependence of the parton distribution function  for weak coupling  in vacuum (dashed-red), strong coupling in
  vacuum (dark-solid-blue) and near a holographic Fermi surface (light-solid-green).}
  \label{fig_pdf}
\end{figure}

\begin{widetext}
\be
\label{pdf}
\left(\frac 1{q^2}\right)^{\tau-1}\,x_k^{\tau+1}(1-x_k)^{\tau-2}\rightarrow \left(\frac 1{q^2}\right)^{\nu_k+1}
x_k^{\nu_k+2}(1-x_k)^{\tau-\frac 32}\,{}_2F_1^2\left(\tau_+, \tau_-, \tau-1, 1-x_k\right)\nonumber\\
\ee
\end{widetext}
with $2\tau_\pm=\tau\pm (\nu_k+1/2)$ and the twist parameter $\tau=mR+3/2$.
Near the black hole horizon, the parton distribution function develops a modified scaling law, but it is still seen to
 vanish at the end points $x_k=0,1$. In Fig.~\ref{fig_pdf} we show the modified behavior
of the partonic distribution function in (\ref{pdf}) for fixed $q^2$, $\tau=3$ and $\nu_k=\frac 12$ versus $x_k$ as the  light-solid
curve (green). The comparison is with the large-x dependence of the nucleon for weak coupling dashed curve (red),
and strong coupling dark-solid curve (blue). Near the black hole horizon, the distribution function is shifted to intermediate-x.
With our choice of parameters, the holographic result (\ref{pdf}) reduces to $x_k^{\frac 12}(1-x_k)^{\frac 32}$, in comparison
to the strong coupling result in vacuum $x_k^4(1-x_k)^2$, and the weak coupling result also in the vacuum
$x_k^{\frac 12}(1-x_k)^{3}$. Remarkably, the formation of a holographic fermionic surface
through the  AdS$_2\times $R$^3$ reduction, is to shift the
holographic partonic distribution to intermediate-x, and modify the hard scattering rule.

For our choice of DIS kinematics, the non-normalized
$\tilde F_2$ structure function (\ref{A2}) follows from (\ref{GRxx5}) in the form ($q_0\approx q_x$)

\be
 \label{GRxxx5}
\tilde F_2(x_A, q)=\frac{4}{\pi}x_A^3\frac{E_A^2}{q^2}\,{\rm Im}\,{\tilde G}_{xx}^F(q)
\ee
with again $Q^2=q^2>0$.
Modulo the dispersion relation and the anomalous exponents that characterize the holographic fermions in the reduced
AdS$_2\times$R$^3$ geometry, the results (\ref{GRxx5}) and (\ref{GRxxx5}) are similar to the ones we derived recently
in~\cite{Mamo:2018eoy} using general arguments.

\begin{figure}[!htb]
 \includegraphics[height=10cm]{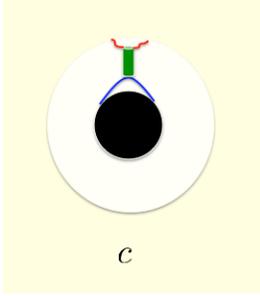}
  \caption{Absorbtive part of the R-current  on a nucleus as an extremal RN-AdS black hole:
  (c) fermionic contribution to order $N_c^0$ due to a closed string exchange in bulk at low-x.}
  \label{fig_low}
\end{figure}

\section{Fermionic contribution at low-x}

In the DIS regime with $q\gg Q^2$ or low-x, the structure functions are dominated by the exchange of a Pomeron,
a multigluon exchange with vacuum quantum numbers. In holography, this exchange
is described either through a closed surface exchange~\cite{SURFACE} or a graviton~\cite{GRAVITON} in bulk.
For the latter, this regime was identified in the range $e^{-\sqrt{\lambda}}\ll x\ll 1/\sqrt{\lambda}$ where the
exchange involves the string scattering amplitude. Since $x\gg e^{-\sqrt{\lambda}}$, the strings are small compared
to the size of the AdS space so that the scattering amplitude is quasi-local with almost flat space signature.



\subsection{General set up}

The 10-dimensional tree-level effective actionthat describes the scattering of an R-photon off bulk
quantum fermions at low-x  reads~\cite{GRAVITON}

\begin{widetext}
\be
\label{SPOM}
{\cal S}=&&\int d^{10}x\,\sqrt{-g_{10}}\left({\cal K V}\right)_{t=0}\nonumber\\
=&&\frac i8\int d^{10}x\,\sqrt{-g_{10}}\left(4v^av_a\,\bar{\psi}\Gamma_m\partial^p\psi-g^p_{m}
\left(\bar\psi\Gamma^M \partial_M\psi\,v_av^a+2v_av_b\bar\psi\Gamma^a\partial^b\psi\right)
\right)\,F^{mn}F_{pn}\,{\cal V}|_{t=0}
\ee
\end{widetext}
where $v^a$ are the Killing vectors for the compact  part of the 10-dimensional space.
The forward R-current scattering amplitude follows from pertinent variation with respect to the R-field.
Here ${\cal K}$ refers to the kinematical factor involving the fermions $\psi$ and the R-field
strength $F$, and ${\cal V}$ is the exchanged flat space  10-dimensional Virazorro-Shapiro string  amplitude

\be
\label{VEN}
{\cal V}=\frac {\alpha^{\prime 3}{\tilde s}^2}{64}\prod_{\tilde\beta=\tilde s,\tilde t , \tilde u}
\frac{\Gamma\left(-\frac{\alpha^\prime\tilde \beta}4\right)}{\Gamma\left(1+\frac{\alpha^\prime\tilde \beta}4\right)}
\ee
as illustrated in Fig.~\ref{fig_low}.
The 10-dimensional Mandelstam variables $\tilde s,\tilde t$ are related to the
4-dimensional ones $s,t$ through

\be
\alpha^\prime \tilde s=&&  \alpha^\prime s \frac{z^2}{R^2} + {\cal O}\left(\frac 1{\sqrt{\lambda}}\right)\nonumber\\
\alpha^\prime \tilde t=&&   \alpha^\prime  t \frac{z^2}{R^2}+{\cal O}\left(\frac 1{\sqrt{\lambda}}\right)
\ee
with the warping made explicit. 

The imaginary part of the string amplitude (\ref{VEN}) is
\be
{\rm Im}{\cal V}|_{t=0} =\frac {\pi\alpha^\prime}4\sum_{j=1}^\infty\,j^{{\cal O}\left(\frac 1{\sqrt{\lambda}}\right)}\,
\delta\left(j -\frac{\alpha^\prime\tilde s}4\right)
\ee
with the delta-function summing over the closed string Regge trajectory.
At low-x we have $s\sim 1/x$ and $j\sim s\sim 1/x$,
so that for ${\rm ln}(1/x)/\sqrt{\lambda}\ll 1$,
\be
j^{{\cal O}\left(\frac 1{\sqrt{\lambda}}\right)}\sim \left(\frac 1x\right)^{{\cal O}\left(\frac 1{\sqrt{\lambda}}\right)}\sim 1
\ee

We now recall that the field strength $F^{mn}$ describes the bulk-to-boundary R-field-strength with incoming momentum
$q^\mu$ and outgoing momentum $q^{\mu}$, while $\psi$ describes the bulk fermion with incoming
and outgoing momentum $k^\mu$ in the anomalous Fermi surface. The low-x regime with $x\ll 1$ corresponds to the
kinematical regime $q\cdot k\gg q^2\gg k^2$, so that the dominant contribution in ${\cal K}$ is
the term with the spin contraction of the form $(q\cdot k)$, i.e.  the first term in (\ref{SPOM}). Using

\be
\label{X1X}
\psi(k)\rightarrow\psi(k)\times{\mathbb Y}(v)
\ee
and normalizing
\be
\label{X2X}
\int_{S^5} d\hat v\,\sqrt{g_{S^5}}\,v^av_a\,|{\mathbb Y}(v)|^2=c_5R^2
\ee
where ${\mathbb Y}(v)$ is a spherical harmonic in $S^5$ in (\ref{SPOM}), we can write down the one-loop effective action ${\cal S}_F$ for the diagram shown in Fig.~\ref{fig_low} as

\begin{widetext}
\be
\label{SIM}
{\cal S}_F[A_{\mu}^{(0)}\equiv n_{\mu}]=\frac{i}2c_{5}R^2\,\int \frac{d^{4} q}{(2 \pi)^{4}}\int \frac{d^{4} k}{(2 \pi)^{4}}\,\int dr\,\sqrt{-g}\,
F^{\mu m}(-q)F_{\nu  m}(q)\,{\rm Tr}\left(\,D_F(r,r, k)\Gamma_{\mu}(-ik)^\nu\right)\,{\cal V}_F|_{t=0}\nonumber\\
\ee
\end{widetext}

We now choose  the polarizations to be transverse with the additional axial gauge condition  $a_{r}=0$,
 so that the boundary-to-bulk R-field is

\be
\label{X3X}
a_\mu( r, \vec q)=\left(\frac{R^2}rK_1\left(\frac {qR^2}r\right)\right)\,n_\mu(q)\,e^{iq\cdot x}\,
\ee
The corresponding field strengths are

\begin{eqnarray}\label{X4X}
&&F_{\mu\nu}(q) = i(q_\mu n_\nu - n_\mu q_\nu) \frac{q R^2}{r} K_1\left(\frac{qR^2}r\right)
e^{i q\cdot x}\ ,
\nonumber\\
&&F_{\mu r}(q) = n_\mu q^2 \frac{R^4}{r^3} K_0\left(\frac{qR^2}r\right) e^{iq\cdot x}\nonumber\\
\end{eqnarray}
and their contraction is

\begin{eqnarray}\label{X5X}
F_{\mu p} F_\nu{}^p =&&+n_\mu n_\nu \frac{q^4R^6}{r^4} \left(K_0^2\left(\frac{qR^2}r\right) + K_1^2\left(\frac{qR^2}r\right) \right)\nonumber\\
&&+{q_\mu q_\nu}n^2 \frac{q^2R^6}{r^4} K_1^2\left(\frac{qR^2}r\right)\nonumber\\
\end{eqnarray}


\subsection{Low-x near the horizon}

To analyze the  low-x contribution of the fermions near the horizon, we will focus on the graviton exchange
and make use of warped momenta $\tilde q$ throughout this section.
For small energy transfer $\tilde{q}_0\ll\mu$, the bulk-to-bulk propagator for transverse graviton $h_x^{~y}(\tilde{q}_0,\tilde{q}_x)$ can be written as

\be
&&G_{xy,xy}(\tilde{q}_0,\tilde{q}_x,r_1,r_2)=\nonumber\\
&&\phi(\tilde{q}_0,\tilde{q}_x,r_1)G^B_{xy,xy}(\tilde{q}_0,\tilde{q}_x)\phi(q_0,q_x,r_2)
\ee
where $\phi(\tilde{q}_0,\tilde{q}_x,r_1)$ is the normalizable wave function of the graviton, and $G^B_{xy,xy}$ is its boundary Green's function
\be
\label{GXYXY}
G^B_{xy,xy}(\tilde{q}_0,\tilde{q}_x)=\tilde{q}_0^2\,G(\tilde{q}_0,\tilde{q}_x)
\ee
where ${\rm Re}\,G(\tilde{q}_0,\tilde{q}_x)=f(\tilde{q}_x,\mu)$ which can be determined from the low-frequency expansion, and \cite{Edalati:2010hk}
\bea
{\rm Im}\,G(\tilde{q}_0,\tilde{q}_x)&&=\frac{3C}{\left({1+\frac{\tilde{q}_0^2}{\mu^2}}\right)^{\frac 12}}
\Big(1+\Big({1+\frac{\tilde{q}_0^2}{\mu^2}}\Big)^{\frac 12}\Big)\nonumber\\&&\times e_0\Big(\frac{\tilde{q}_{x}}{\mu}\Big)
{\rm Im}\mathcal{G}_{\pm}\Big(\frac{\tilde{q}_0}{\mu},\frac{\tilde{q}_x}{\mu}\Big)
\eea
where $C$ is a proportionality constant, $e_0\Big(\frac{\tilde{q}_{x}}{\mu}\Big)$ is a function to be determined from the low-frequency expansion coefficients, and
\be
&&\mathcal{G}_{\pm}\Big(\frac{\tilde{q}_0}{\mu},\frac{\tilde{q}_x}{\mu}\Big)=-2\nu_{\pm}e^{-i\pi\nu_{\pm}}\frac{\Gamma(1-\nu_{\pm})}{1+\nu_{\pm}}\Big(\frac{1}{2}\frac{\tilde{q}_0}{\mu}\Big)^{2\nu_{\pm}}\,,\nonumber\\
\ee
where
\be
\nu_{\pm}=\frac{1}{2}\Big({5+2\frac{\tilde{q}_x^2}{\mu^2}\pm 4\Big({1+\frac{\tilde{q}_x^2}{\mu^2}}\Big)^{\frac 12}}\Big)^{\frac 12}\,.
\ee

Note that for zero energy and momentum transfer ($\tilde{q}_0=0$ and $\tilde{q}_x=0$), the bulk-to-bulk propagator of the graviton
exchange vanishes ${\rm Im}\,G_{xy,xy}(0,0,r_1,r_2)=0$ since $\mathcal{G}_{\pm}(0,0)=0$. Therefore, the t-channel contribution of the graviton  for the current-current correlation function or forward deeply virtual Compton scattering away from the probe limit vanishes. Its Reggeized form
through higher spins (closed string exchange), vanishes as well.

\section{R-ratio for  a quantum corrected RN-AdS black-hole}

\subsection{Particle and energy density at the horizon}

Having assessed the structure functions both at large-x and small-x near the black hole horizon, we now need
to normalize them.  For that we need
to evaluate the contribution of the bulk fermions near the horizon to the particle and energy densities, much like
we did in the probe limit. More specifically, we define

\be
ne_R=&&\left<J^t\right>(qz\ll 1)\nonumber\\
\epsilon=&&\left<T^{tt}\right>(qz\ll 1)
\ee
as the boundary expectation values of the time-component of the R-current and the energy momentum tensor.
The expectation values  follow from the holographic correspondence  in the tadpole approximation in AdS as

\bea
\left<J^t\right>=&&-ie_R\int\frac{d^3k}{(2\pi)^3}\int \frac{dk^0}{2\pi}\,I_K(qz\ll 1)\,
\nonumber\\&&\times \overline{\psi_1} (k)\Gamma^{\underline{t}}\psi_1 (k)
\,{\rm Im Tr}\,{\cal G}_R^{11}(k^0, \vec k)\nonumber\\
\left<T^{tt}\right>=&&\int\frac{d^3k}{(2\pi)^3}\int \frac{dk^0}{2\pi}\,I^\prime_K (qz\ll 1)\,\nonumber\\
&&\times \overline{\psi_1} (k)\Gamma^{\underline{t}}\psi_1 (k)(-ik^0)\,{\rm Im Tr}\,{\cal G}_R^{11}(k^0, \vec k)\nonumber\\
\eea
with $I^\prime_K(qz\ll 1)\approx I_K(qz\ll 1)$  playing the role of a spectral weight, and defined in (\ref{IKZQ}).
Evaluating the momentum integral near the Fermi surface, we find

\bea
n=\frac{\left<J^t\right>}{e_R}&&\approx
\,C_{J}\,h_1C_\theta\frac{ k_F^3(1-\frac{k_R}{k_F})}{\left|\frac 1{v_F}+{\rm Re}\,\Pi^\prime\right|}\nonumber\\
\eea
with  $C_\theta=\frac 1{2\pi}$ and the dimensionless constant

\be
C_{J}=&&R^4(mR_2+\nu_{k_F})\Big(\frac{R_2}{R}k_Fz_{-}+e_R\sqrt{\frac{\tilde{\alpha}}{3}}\Big)\nonumber\\
&&\times \frac{a_+^2(k_0,k)}{(2\nu_k+1)W^2}(2\sqrt{3})^{-2\nu_{k_F}}
\ee
Since $I_K=I_K^\prime$, we have $\epsilon=nk^0$.  Note that $k_0$ is the solution of the transcendental equation~(\ref{transcendental}), which near the Fermi surface $k\rightarrow k_F$ can be solved as $k_0\sim {C_0}/{z_{-}}$  with the dimensionless constants

\be
C_0(\nu_{k_F})\equiv(v_F \tilde{h}_2{\rm Im}\,(e^{i\gamma_F}(-1)^{2\nu_{k_F}}))^{\frac{1}{1-2\nu_{k_F}}}
\ee
and $\tilde{h}_2=z_{-}^{1-2\nu_{k_F}}h_2$. Therefore, for the dense limit near the horizon, we make  the identification
$E_{A}\equiv V_A\,\epsilon=A\,{\epsilon}/{n}=A\,k^0$.

\subsection{Normalized structure functions :\newline  dense regime}

Having determined $n, \epsilon$ in the dense limit near the horizon, we can now normalize the
corresponding structure function (\ref{GRxx5}) through the substitution

\be
\label{TRANS}
{\rm Im}\,\tilde{G}^F_{xx}(q)\rightarrow 2E_A V_A\times {\rm Im}\tilde{G}^F_{xx}(q)=2E_A\,A\frac{{\rm Im}\,\tilde{G}^F_{xx}(q)}{n}\nonumber\\
\ee
The integral in (\ref{GRxx5}) can be evaluated near the fermi surface $k\rightarrow k_F$  with the result

\begin{widetext}
\be \label{GRxx51}
{\rm Im}\,\tilde{G}^F_{xx}(q)\approx \frac 12 \tilde{C}_G(\nu_{k_F})a_+(k_0,k_{\perp}=0)^2\frac {k_{F}^3\Big(1-\frac{k_{R}}{k_{F}}\Big)}{\left|\frac 1{v_F}+{\rm Re}\,\Pi^\prime\right|}
\left(\frac 1{q^2z_{-}^2}\right)^{\nu_{k_F}+1}
x_{k_F}^{\nu_{k_F}+2}(1-x_{k_F})^{\tau-\frac 32}\,{}_2F_1^2\left(\tau_+, \tau_-, \tau-1, 1-x_{k_F}\right)\,,\nonumber\\
\ee
\end{widetext}
Here $\tilde{C}_G(\nu_{k_F})$ and $a_+(k_0,k_{\perp}=0)\approx \overline{c}_2z_{-}\,k_0$ are both dimensionless constants, and $x_{k_{F}}\approx\frac{E_A}{k_0}x_A=\frac{m_N}{k_0}x$ since $\frac{x_A}{x}=\frac{m_N}{E_A}$.  Here $k_0\sim C_0/z_-$ plays the role of the Fermi energy of the quasi-particles in the holographic Fermi liquid near the horizon $z_-$. We recall that

 \be
 \frac{1}{z_{-}}=\frac{r_{-}}{R^2}=\frac{1}{2\sqrt{3}}\,\frac{\mu}{\sqrt{\tilde{\alpha}}}=\frac{R_2}{R}\,\frac{\mu}{\sqrt{\tilde{\alpha}}}\, ,
 \ee
with $\tilde{\alpha}=1$ for a U(1) R-charge,  and
$\tilde{\alpha}=\frac{1}{4}\frac{N_c}{N_f}$ for a D3-D7  U(1) vector charge.

Using \ref{GRxxx5}) together with (\ref{TRANS}-\ref{GRxx51}), we can extract the normalized structure function of the holographic Fermi liquid
($x_{k_F}k_0=xm_N$)

\begin{widetext}
\bea \label{F2ads2}
&&\frac{F_2(x_A,q^2)}{A}=e_R^2C_{AdS2}(e_R,\tau,\tilde{\alpha},v_F,\tilde{h}_2)\left(\frac{\mu^2}{q^2}\right)^{\nu_{k_F}+2}\,x_{k_F}^{\nu_{k_F}+5}(1-x_{k_F})^{\tau-\frac 32}\,{}_2F_1^2\left(\tau_+, \tau_-, \tau-1, 1-x_{k_F}\right)\,,\nonumber\\
\eea
\end{widetext}
where we defined the dimensionless constant
\begin{widetext}
\be
&&C_{AdS2}(e_R,\tau,\tilde{\alpha},v_F,\tilde{h}_2)\equiv
\nonumber\\
&&\Big(\frac{1}{3\tilde{\alpha}}\Big)^{\nu_{k_F}+2}\frac{1}{8}C_{0}^3\frac{(2\nu_{k_F}+1)\Big(\frac{1}{2\sqrt{3}}\tau+\nu_{k_F}-\frac{\sqrt{3}}{4}\Big)\Gamma(\tau+\nu_{k_F}+\frac{3}{2})^2\Gamma(\tau+\nu_{k_F}-\frac{1}{2})^2}{\Big(\frac{k_F}{\mu}\sqrt{\tilde{\alpha}}+\frac{1}{\sqrt{3}}e_R\sqrt{\tilde{\alpha}}\Big)\Gamma(\tau-1)^2}\,, \nonumber\\
\ee
\end{widetext}
with

\be
\label{PARAX}
&&\nu_{k_F}(e_R,\tau,\tilde{\alpha})=
\Big({\frac{1}{12}\Big(\tau-\frac{3}{2}\Big)^2+\frac{k_F^2}{\mu^2}\tilde{\alpha}-\frac{1}{3}e_R^2\tilde{\alpha}}\Big)^{\frac 12}
\nonumber\\
&&C_{0}=\Big(v_{F}\tilde{h}_2\sin(\gamma_F+2\pi\nu_{k_F})\Big)^{\frac{1}{1-2\nu_{k_F}}}\nonumber\\
&&\gamma_{F}={\rm arg}\Big(\Gamma(-2\nu_{k_F})\Big(e^{-2\pi i\nu_{k_F}}-e^{-\frac{2\pi}{\sqrt{3}}e_R\sqrt{\tilde{\alpha}}}\Big)\Big)\nonumber\\
\ee
Note that for a large effective charge $e_R\sqrt{\tilde{\alpha}}\rightarrow \infty$, we have $\gamma_F=-2\pi\nu_{k_F}$ and  $C_0\rightarrow 0$ which implies that the structure function \ref{F2ads2} vanishes in the probe limit $\tilde{\alpha}\sim\frac{N_c}{N_f}\rightarrow \infty$, which is also the regime where the backreaction from the flavor branes can be ignored.

\subsection{R-ratio in the dense regime}

The R-ratio for the quantum corrected black hole consists of:
1/ the leading classical contributions (\ref{F2BH}) plus (\ref{F2BHLX}) both at small and large $x$ respectively;
2/ the sub-leading quantum correction from the emergent Fermi surface (\ref{GLX2}) in the AdS$_2\times$R$^3$ in
the near horizon approximation; 3/ the sub-leading quantum correction from a normal Fermi liquid  far from the
horizon in the dilute or probe approximation  (\ref{f2dilute}) to be detailed below.  To map this ratio on that of a
dense nucleus, we follow~\cite{Mamo:2018ync} and define

\be \label{rdense}
R_{\rm dense}(x,q^2)\equiv
\frac{\frac 1A F_2^{\rm dense}(x,q^2)}{F_2^{\rm nucleon}(x,q^2)}
\ee
with the dense structure function

\begin{widetext}
\bea \label{f2dense}
\frac{F_2^{\rm dense}(x,q^2)}{A}&&\approx\bigg( C_T\left(\frac{3q^2}{4m_N^2}\right)^{\frac 23}\,x^{\frac 13}
+\frac{3C_L}{2x}\left(\frac{3x^2q^2}{4m_N^2}\right)D(x)\bigg)\nonumber\\
&&+\bigg(C_{AdS2}\,e_R^2\,\left(\frac{\mu^2}{q^2}\right)^{\nu_{k_F}+2}\,x_{k_F}^{\nu_{k_F}+5}(1-x_{k_F})^{\tau-\frac 32}\,{}_2F_1^2\left(\tau_+, \tau_-, \tau-1, 1-x_{k_F}\right)
+\mathbb C_{1}\frac{F_2^{\rm AdS5}(x,q^2)}{A}\bigg)\,,\nonumber\\
\eea
\end{widetext}
normalized by the nucleon structure function (\ref{f2nucleon}). 
For clarity, we recap the different definitions used for parton-x ($0\leq x_A\leq 1$)  and entering (\ref{f2dense}) and the normalization
(\ref{f2nucleon})

\bea
&&x_AE_A=xm_N\nonumber\\
&&xm_N\approx x_{k_F}k_0\equiv x_{k_F} k_F
\eea
$x_A$ refers to the parton fraction in a nucleus, $x$ refers to the parton fraction in a nucleon within a nucleus,
and  $x_{k_F}$ refers to the parton fraction in relation to $k_0\sim C_0/z_-$ from the emergent Fermi energy.

In (\ref{f2dense}), the first bracket is the leading and classical contribution, and the second bracket
is the subleading and quantum contribution. More specifically,
the first (\ref{F2BH}) (low-x) and second (\ref{F2BHLX}) (large-x) contributions stem from DIS scattering on the bulk classical black hole.
The third (\ref{GLX2}) contribution stems from DIS scattering off the emerging holographic liquid close to the black hole horizon. The fourth and last
contribution $\frac{F_2^{\rm AdS5}(x,q^2)}{A}\equiv\frac{F_2^{\rm dilute}(x,q^2)}{A}$ (\ref{f2dilute}) stems from DIS scattering off the distorted Fermi liquid far from the horizon in the probe approximation to be discussed thoroughly below.
The quantum correction  near the black hole is vanishingly small at small-x.


The relative and arbitrary normalization $\mathbb C_{1}=0.07$ is introduced to account for the normal Fermi liquid contribution far from the horizon which is asymptotically
AdS$_5$  as we discussed earlier. It will be estimated in the dilute limit below and added  to the near horizon contribution. A more quantitative calculation using the exact response function throughout the holographic space distorted by the RN-AdS extremal black hole, that reduces to AdS$_2$ near the horizon and asymptotes the dilute limit near the boundary,  is numerically intensive  and goes outside the scope of this work. We note that the values of $\mathbb C_1\leq 0.07$ keep the AdS$_2$ plus AdS$_5$
quantum corrections subleading in comparison to the leading RN-AdS black hole contribution for all values of $k_F$ and most parton-x.

To quantify each of the contributions in the R-ratio, we now need to fix the parameters entering this expression, many of which are tied by holography.
We first fix the explicit holographic parameters: $\tilde\alpha=N_c/4N_f=1$ (ratio of branes), $2\pi^2c_5/\sqrt{4\pi\lambda}=0.01$
(strong coupling) and $e_R=0.3$ (charge of the probe fermions). Next, we fix the scaling parameters entering in the nucleon pdf:
$\tau=3$ (hard scaling law) and $j=0.08$ (Pomeron intercept). The nucleon confining scale enters through $\beta=1/(m_Nz_-)^2\rightarrow \tilde \beta$
asymptotically (\ref{f2nucleon}).  We fix it to $\tilde\beta=17.65$. 
Finally, we fix the parameters of the emergent Fermi surface: $v_F=1$ (Fermi velocity),  $\tilde{h}_2=1$ for simplicity,
$\mu/m_N=1.2$ (chemical potential) for a typical nucleus

\begin{widetext}

\begin{figure}[!htb]
 \includegraphics[height=5cm]{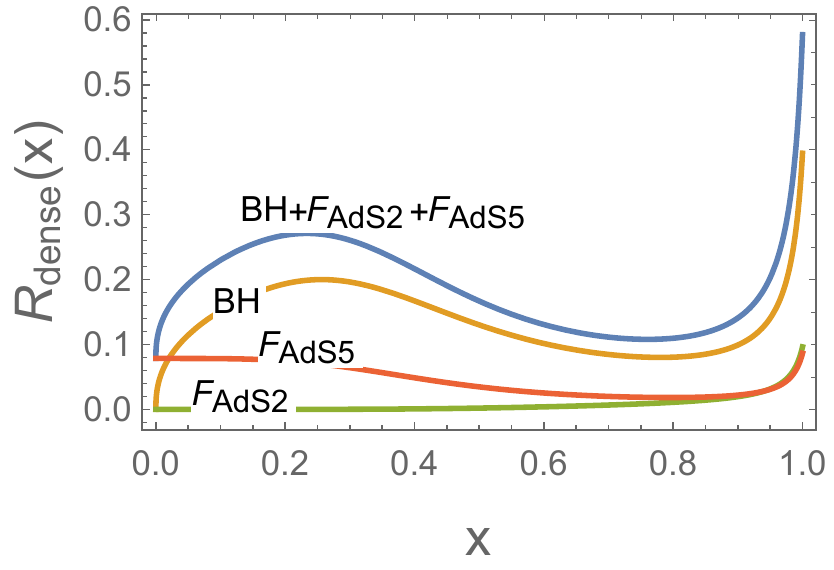}
  \caption{Dense R-ratio (\ref{rdense}) with $k_F/m_N=4$. See text.}
\label{dense-ratio1}
\end{figure}

\begin{figure}[!htb]
 \includegraphics[height=5cm]{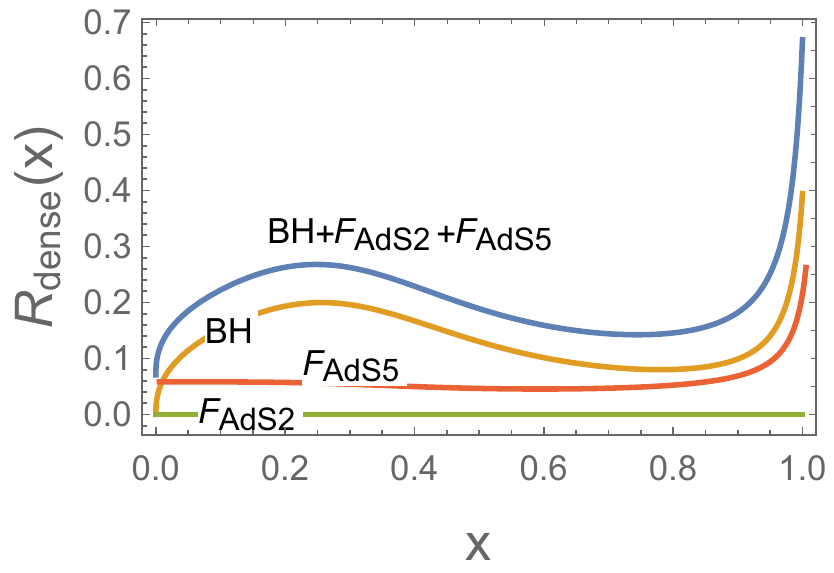}
 \caption{Dense R-ratio (\ref{rdense}) with $k_F/m_N=0.5$. See text.}
\label{dense-ratio2}
\end{figure}

\begin{figure}[!htb]
 \includegraphics[height=5cm]{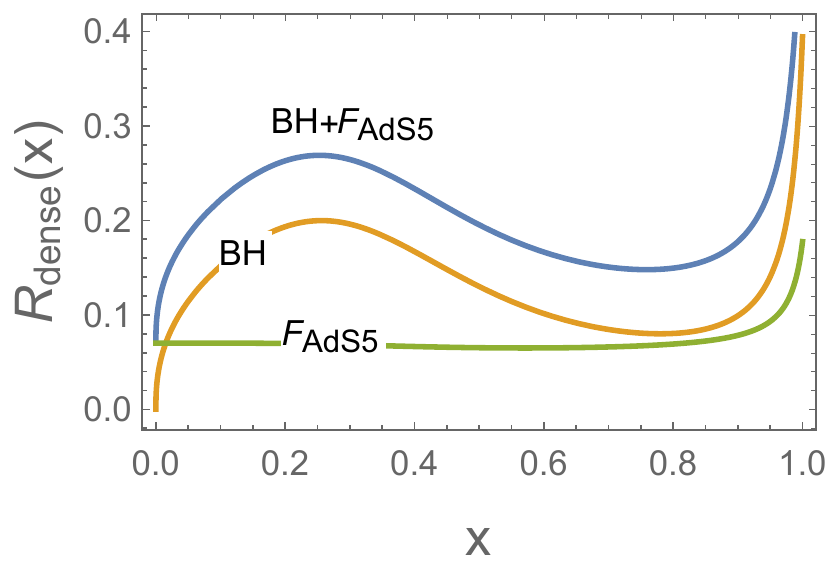}
 \caption{Dense R-ratio (\ref{rdense}) with $k_F/m_N=0.3$, and without the contribution from AdS$_2$. See text.}
\label{dense-ratio3}
\end{figure}

\end{widetext}

With these parameters fixed, we show in Figs.~\ref{dense-ratio1}-\ref{dense-ratio3}, the dense R-ratio versus x for different Fermi momenta $k_F/m_N=4,0.5,0.3$.
For  $k_F/m_N\geq 0.4\approx k_R/m_N$ the contribution from the emergent AdS$_2$ Fermi surface near the horizon is small but real. This contribution disappears
for $k_F<k_R$ as we noted in (\ref{k2kR}), and only the AdS$_5$ contribution far from the horizon remains. The contribution from the emergent AdS$_2$ Fermi surface
becomes comparable to the far horizon AdS$_5$ contribution at large-x and only for large values of $k_F/m_N$. For all values of  $k_F/m_N$ and most values of $x$,
the leading black hole contribution is dominant.  Some of the  key features of DIS scattering on a nucleus are already captured by Figs.~\ref{dense-ratio1}-\ref{dense-ratio3}
with shadowing and anti-shadowing at small-x due to the coherent scattering on the black-hole, and Fermi motion at large-x that is increasingly
apparent from DIS scattering on the AdS$_5$ from the far horizon part at small $k_F/m_N$. We will return to these important physical issues
below through a more realistic model for DIS scattering on a finite nucleus with comparison to the existing world-data from  DIS scattering on  light and heavy nuclei.

\section{DIS in the Probe limit :\newline dilute  regime}

 We now consider  scattering in the probe limit, where the bulk fermions carry a density
without affecting the underlying AdS$_5$ geometry (with or without a wall), i.e.  $\frac{\mu}{\sqrt{\tilde{\alpha}}}\rightarrow 0$
with $\frac{\mu}{\sqrt{\tilde{\alpha}}}\times e_R\sqrt{\tilde{\alpha}}=\mu_e$ fixed where $\tilde{\alpha}\sim\frac{N_c}{N_f}\gg 1$ and $\mu$ is the chemical potential.
This limit, can be regarded as the UV completion for the emergent Fermi surface in bulk, but also can be considered as DIS scattering on a very dilute nucleus. This
amounts to using the free spectral form (\ref{specBB2}) with the substitution

\be
{\rm Im}\,{\cal G}_F^{\alpha\gamma}(\omega_1, k)\rightarrow \pi n_F(\omega_1, \vec k)\,\delta(k^2+\omega_1^2)\,\delta^{\alpha\gamma}
\label{nfermion}
\ee
Here  $n_F(\omega_1,\vec k)=\theta(\mu_e-(k^2+\omega_1^2)^{\frac 12})$ is the Fermi occupation factor for a fermion of momentum $\vec k$,
mass $\omega_1$  and Fermi energy $\mu_e$, and the vacuum (AdS$_5$) wavefunctions. For the confining
case, the mass $\omega_1$ is quantized.
This analysis complements the one we have discussed recently using generic arguments based on a density
expansion of a trapped Fermi liquid~\cite{Mamo:2018eoy}.

\subsection{Large-x}

With this in mind,  consider the case of scattering in the ultraviolet region of the black-hole, with the
hard fermion of momentum $k+q$ and the remaining fermion of momentum $k$
treated in the probe approximation. This example will help clarify the relationship between our analysis and that in
\cite{Polchinski:2002jw}. For that we use the vacuum propagator (\ref{spectrep2}) for both the hard
fermion and the density modified propagator (\ref{spectrep}) and (\ref{nfermion})
for the soft fermion  in (\ref{GRx3}),

\begin{widetext}
 \be
{\rm Im}\,\tilde{G}_F^{\mu\nu}(q) &&=
\int \frac{d^{4} k}{(2 \pi)^{4}}
 \int d \om_1 \omega_1\int d \om_2 \omega_2
{\rm Re}\,{\rm Tr}\,\Big({\mathcal{G}}_F (\om_1,k+q)\, \Lambda^\mu (\om_1;q;\om_2) \,\mathcal{G}_{F}(\om_2,k)  \,
 \Lambda^\nu (\om_2;q;\om_1)\Big)\,, \nonumber\\
 &&=\int \frac{d^{3} k}{(2 \pi)^{3}}
 \int_{-|q^0|}^{0}\frac{dk^{0}}{2\pi}\int d \om_1 \omega_1\int d \om_2\omega_2
\,{\rm Tr}\,\Big({\rm Im}\,{\mathcal{G}}_R(\om_1,k+q)\,\Lambda^\mu (\om_1;q;\om_2)\,{\rm Im}\,\mathcal{G}_{R}(\om_2,k)\,\Lambda^\nu (\om_2;q;\om_1) \Big)\, \nonumber\\
 \label{GRxwt}
\ee
\end{widetext}
where the hard vertices are defined as

\bea
\label{vertex11wt}
 &&\Lambda^\mu (\om_1;q;\om_2)= \nonumber \\
 &&  \int dr_{2} \sqrt{g(r_{2})} \,\overline{\psinorm} (r_2,\om_1) \, Q^{\mu} (r_{2};q) \, \psinorm (r_2,\om_2)\,,\nonumber\\
 &&\Lambda^\nu (\om_2;q;\om_1)= \nonumber \\
 &&  \int dr_{1} \sqrt{g(r_{1})} \,\overline{\psinorm} (r_{1},\om_2) \, Q^{\nu} (r_{1};q) \,
 \psinorm (r_1,\omega_1)\,,\nonumber\\
\eea
Here  $\psinorm (r,\omega)$ and $\overline\psinorm (r,\omega)$ are the hard  wave functions~\cite{Polchinski:2002jw}

\be \label{vacwf}
&&\psinorm (z,\omega)=z^{5/2}\left[
J_{mR-1/2}(\omega z)P_{+}+J_{mR+1/2}(\omega z)P_{-}\right]\,,\nonumber\\
&&\overline{\psinorm}(z,\omega)=z^{5/2}\left[
P_{-}J_{mR-1/2}(\omega z)+J_{mR+1/2}(\omega z)P_{+}\right]\,.\nonumber\\
\ee
with the chiral projectors $P_{\pm}=\frac{1}{2}(1\pm \gamma_{5})$.
In this regime, the imaginary parts are reducible to on-shell delta functions

\be
\label{A6}
&&{\rm Im}\,\mathcal{G}_R(\om_1,k+q)\, {\rm Im}\,\mathcal{G}_{R}(\om_2,k^0,\vk)\rightarrow\nonumber\\
&&(-(k\hspace{-6pt}\slash+q\hspace{-6pt}\slash)+\omega_1)\pi\,\delta((k+q)^2+\omega_1^2)\nonumber\\
&&\times(-k\hspace{-6pt}\slash+\omega_2)\pi\,n_F(\omega_2, \vec k)\,\delta(k^2+\om_2^2)\nonumber
\ee
after using (\ref{nfermion}). Recall that $n_F(\omega, \vec k)$ is the Fermi disribution for a fermion
of mass $\omega$ and momentum $\vec k$ near the boundary in the probe limit.
With this in mind,  (\ref{GRxwt}) becomes

\begin{widetext}
 \be
{\rm Im}\,\tilde{G}_F^{\mu\nu}(q) &&\approx \int \frac{d^{3} k}{(2 \pi)^{3}}
 \int_{-|q^0|}^{0}\frac{dk^{0}}{2\pi}\int_{0}^{q^2}\frac{d\om_2^2}{2}\int_{0}^{(q-\om_2)^2}\frac{d\om_1^2}{2} \nonumber\\
&&\times \tr \Big((-(k\hspace{-6pt}\slash+q\hspace{-6pt}\slash)+\omega_1)\,\Lambda^\mu (\om_1;q;\om_2)\,(-k\hspace{-6pt}\slash+\omega_2)\Lambda^\nu (\om_2;q;\om_1)\Big)\,\pi\,\delta((k+q)^2+\omega_1^2)\pi\,\delta(k^2+\om_2^2)\,, \nonumber\\
&&=\int \frac{d^{3} k}{(2 \pi)^{3}}
 \int_{-|q^0|}^{0}\frac{dk^{0}}{2\pi}\int_{0}^{q^2}\frac{d\om_2^2}{2}\int_{0}^{(q-\om_2)^2} \frac{d\om_1^2}{2}I_{zv}^2(\om_1,\om_2,q)\nonumber\\
&&\times \tr \Big((-(k\hspace{-6pt}\slash+q\hspace{-6pt}\slash)+\omega_1)\,\gamma^\mu \,(-k\hspace{-6pt}\slash+\omega_2)\gamma^\nu P_{+}\Big)\,\pi\,\delta((k+q)^2+\omega_1^2)\pi n_F(\omega_2, \vec k)\,\delta(k^2+\om_2^2)\,, \nonumber\\
 \label{GRxwt2}
\ee
\end{widetext}
with the physical condition $\om_1+\om_2 < q$ (i.e., a meson or virtual photon of mass $q$ decaying into KK-fermions of masses $\om_1$ and $\om_2$), and
\begin{widetext}
\be
\label{Iv}
I_{zv}(\om_1,\om_2,q)&&=e_R\,R^4\,q\,\int_{0}^{\infty}dz\,z^2J_{mR-1/2}(\om_1 z)J_{mR-1/2}(\om_2 z)K_{1}(qz)\nonumber\\
&&\approx e_R\,R^4\frac{2^{-(mR-1/2)}}{\Gamma(mR+1/2)}\om_2^{mR-\frac 12}\,q\,\int_{0}^{\infty}dz\,z^{mR+\frac 32}J_{mR-1/2}(\om_1 z)K_{1}(qz)\nonumber\\
&&\approx 2e_R\,R^4(mR+1/2)\,\frac 1{q^2}\, \left(\frac{\omega_2}{q}\right)^{mR-\frac 12}\,\Big(\frac{\om_1}{q}\Big)^{mR-\frac{1}{2}}\Big(1+\frac{\om_{1}^2}{q^2}\Big)^{-(mR+\frac{3}{2})}\nonumber\\
\ee
\end{widetext}
where  we made use of the approximation

\be
J_{mR-1/2}(\om_2 z)\approx\frac{2^{-(mR-1/2)}}{\Gamma(mR+1/2)}(\om_2z)^{mR-\frac 12}
\ee
 for $\om_2z\ll 1$. Note that without making the approximation $\om_2z\ll 1$, the above integral $I_{z\nu}\sim F_4(a,b;c,d;-\frac{\om_1^2}{q^2},-\frac{\om_2^2}{q^2})$~\cite{Hung:2010pe,Jorrin:2016rbx} where $F_4$ is the fourth Appell series of hypergeometric functions which is indeed convergent only for $\om_1+\om_2<q$.

 The integral in (\ref{Iv}) is in agreement with the R-current scattering on a dilatino in~\cite{Polchinski:2002jw}.
Evaluating the integral in (\ref{GRxwt2}) over $\om_1$ using the delta-function $\delta(\omega_1^2-s_k)$, and using (\ref{Iv}), we
have

\begin{widetext}
 \be
{\rm Im}\,\tilde{G}_F^{\mu\nu}(q) &&\approx \frac{\pi}{2}\int \frac{d^{3} k}{(2 \pi)^{3}}
 \int_{-|q^0|}^{0}\frac{dk^{0}}{2\pi}\int_{0}^{q^2}\frac{d\om_2^2}{2}
\,I_{zv}^2(\sqrt{s_k},\om_2,q){\rm Tr}\,\Big((-(k\hspace{-6pt}\slash+q\hspace{-6pt}\slash)+\sqrt{s_k})\,\gamma^\mu \,(-k\hspace{-6pt}\slash+\omega_2)\gamma^\nu P_{+}\Big)\pi n_F\,\delta(k^2+\om_2^2)\,, \nonumber\\
 \label{GRxwt3}
\ee
\end{widetext}
where $s_k=-(k+q)^2$. The evaluation of the remaining $k^0$-integral in (\ref{GRxwt3}) using the last delta-function,
yields

\begin{widetext}
 \be
{\rm Im}\,\tilde{G}_F^{\mu\nu}(q) &&\approx\frac{\pi}{4} \int_{0}^{q^2}\frac{d\om_2^2}{2}\int \frac{d^{3} k}{(2 \pi)^{3}}
\,I_{zv}^2(\sqrt{s_k},\om_2,q)\,{\rm Tr}\, \Big((-(k\hspace{-6pt}\slash+q\hspace{-6pt}\slash)+\sqrt{s_k})\,\gamma^\mu \,(-k\hspace{-6pt}\slash+\omega_2)\gamma^\nu P_{+}\Big)\,\frac{n_F(\omega_2,\vec k)}{2E_{k}}\,, \nonumber\\
 \label{GRxwt4}
\ee
\end{widetext}
where $E_k=\sqrt{|\vec k|^2+\om_2^2}<|q^0|$, and $k^0=E_{k}$.

To extract the structure functions (\ref{A2}) from (\ref{GRxwt4}) we carry the spin trace by contacting with the time-like
frame vector $n^\mu=(1,0,0,0)$,

\begin{widetext}
 \be
n_{\mu}n_{\nu}{\rm Im}\,\tilde{G}_F^{\mu\nu}(q) &&\approx \frac{\pi}{4} \int_{0}^{q^2}\frac{d\om_2^2}{2}\int \frac{d^{3} k}{(2 \pi)^{3}}
\,I_{zv}^2(\sqrt{s_k},\om_2,q)\tr \Big((-(k\hspace{-6pt}\slash+q\hspace{-6pt}\slash)+\sqrt{s_k})\,n\hspace{-6pt}\slash \,(-k\hspace{-6pt}\slash+\omega_2)n\hspace{-6pt}\slash P_{+}\Big)\,\frac{n_F(\omega_2,\vec k)}{2E_{k}}\,, \nonumber\\
&&\approx \pi   \int_{0}^{q^2}\frac{d\om_2^2}{2}\int \frac{d^{3} k}{(2 \pi)^{3}}
\,I_{zv}^2(\sqrt{s_k},\om_2,q)\,\Big((n\cdot k)^2-\frac{1}{2}(k\cdot q+\,\sqrt{s_k}\,\om_2)\,n^2\Big)\,\frac{n_F(\omega_2,\vec k)}{2E_{k}}\,, \nonumber\\
 \label{GRxwt5}
\ee
\end{widetext}
where we have assumed $n\cdot q\approx 0$ and $k^2\approx 0$. Note that the trace in (\ref{GRxwt5}) is the same trace evaluated in~\cite{Polchinski:2002jw} for $\om_2=0$ (see their Eq.72 ). Using (\ref{A2}) with $x_A=-q^2/2P_A\cdot q$, we can now
 extract the structure functions $F_{1,2}(x_A,q^2)$ of a state with momentum $P_A^\mu$ from (\ref{GRxwt5})

\begin{widetext}
 \be
&&\tilde{F}_{2}(x_A,q^2)\approx 2\pi^2\frac{q^2}{2x_A\,(n\cdot P_{A})^2}   \int_{0}^{\Lambda^2<q^2}\frac{d\om_2^2}{2}\int \frac{d^{3} k}{(2 \pi)^{3}}
\,I_{zv}^2(\sqrt{s_k},\om_2,q)\,(n\cdot k)^2\,\frac{n_F(\omega_2,\vec k)}{2E_{k}}\,, \nonumber\\
&&\tilde{F}_{1}(x_A,q^2)\approx -\pi^2     \int_{0}^{\Lambda^2<q^2}\frac{d\om_2^2}{2}\int \frac{d^{3} k}{(2 \pi)^{3}}
\,I_{zv}^2(\sqrt{s_k},\om_2,q)\,(k\cdot q+\,\sqrt{s_k}\,\om_2)\,\frac{n_F(\omega_2,\vec k)}{2E_{k}}\,, \nonumber\\
 \label{GRxwt6}
\ee
\end{widetext}
We have $x_k=-q^2/2k\cdot q$, $s_{k}=-(k+q)^2\approx-q^2(1-1/x_k)$ and $k^0=E_{k}=(|\vec k|^2+\om_2^2)^{\frac 12}<|q^0|$.
We can reduce $I_{zv}(\sqrt{s_k},\om_2,q)$  in  (\ref{Iv}) in terms of $x_k$ as

\begin{widetext}
\be \label{Ivx}
I_{zv}(\sqrt{s_k},\om_2,q)&&\approx 2e_R\,R^4(mR+1/2)\,\om_2^{mR-1/2}\,q^{-(mR+3/2)}\,x_k^{\frac{1}{2}(mR+7/2)}(1-x_k)^{\frac{1}{2}(mR-1/2)}\,.\nonumber\\
\ee
for the mass range $\omega_2\leq \Lambda$.
Using (\ref{Ivx}), we can re-write the structure functions (\ref{GRxwt6}) in terms of $x_k$ as

 \be
\tilde{F}_{2}(x_A,q^2)&&\approx 2^2\pi^2e_R^2\,R^8(\tau-1)^2\nonumber\\
&&\times\,\Big(\frac{1}{q^2}\Big)^{\tau-1}\,\frac{1}{x_A\,(n\cdot P_{A})^2}   \int_{0}^{\Lambda^2<q^2}\frac{d\om_2^2}{2}\,{(\om_2^2)^{\tau-2}}
\int \frac{d^{3} k}{(2 \pi)^{3}}\,\frac{n_F(\omega_2,\vec k)}{2E_{k}}\,  (n\cdot k)^2
\,x_k^{\tau+2}(1-x_k)^{\tau-2}\, \nonumber\\
\tilde{F}_{1}(x_A,q^2)&&\approx 2\pi^2e_R^2\,R^8(\tau-1)^2\nonumber\\
&&\times\,\Big(\frac{1}{q^2}\Big)^{\tau-1}\,  \int_{0}^{\Lambda^2<q^2}\frac{d\om_2^2}{2}\,(\om_2^2)^{\tau-2}\int \frac{d^{3} k}{(2 \pi)^{3}}
\frac{n_F(\omega_2,\vec k)}{2E_{k}}\,
\,x_k^{\tau+1}(1-x_k)^{\tau-2}\,, \nonumber\\
 \label{GRxwt7}
\ee
\end{widetext}
with the twist parameter is  $\tau=mR+3/2$, following the approximation  ($\om_2\ll q$)

\be
\sqrt{s_k}\,\frac{\om_2}{q^2}\approx \left(\frac 1{x_k}-1\right)^{\frac 12}\frac{\om_2}{q}\approx 0
\ee

\subsection{Low-x}

In contrast to the dense limit in (\ref{GXYXY}),  the bulk-to-bulk graviton propagator in the probe limit, is given by

\be
&&G_{xy,xy}(\tilde{q}_0,\tilde{q}_x,z,z')=\int_0^\infty\,\frac{d\om^2}{2}\frac{z^2J_{\Delta}(z\om)z'^2J_{\Delta}(z'\om)}{-t+\om^2-i\epsilon}\nonumber\\
\ee
where $t=-\tilde{q}^2=\tilde{q}_0^2-\tilde{q}_x^2$. Therefore, $G_{xy,xy}(\tilde{q}_0=0,\tilde{q}_x=0,z,z')\neq 0$ does not vanish
 in the probe limit. In this limit,  the graviton exchange Reggeizes by including  higher spin-j (stringy) exchange as

\be
&&\mathcal{K}(j,\tilde{q}_0,\tilde{q}_x,z,z')=\int_0^\infty\,\frac{d\om^2}{2}\frac{z^2J_{\tilde{\Delta}(j)}(z\om)z'^2J_{\tilde{\Delta}(j)}(z'\om)}{-t+\om^2-i\epsilon}\nonumber\\
\ee
with $\tilde{\Delta}(j)=\Big({4+2\sqrt{\lambda}(j-2)}\Big)^{\frac 12}$.


With this in mind, we now consider the case of the one-loop fermionic contribution in the probe limit at low-x.
In this regime, the bulk-to-bulk fermion propagator is of the form~\cite{Gao:2009se}

\be
\label{feynman2}
D_F(r,r^\prime ,k)=-\int \dfrac{d\om^2}{2}\psinorm(r)\frac{-ik\hspace{-6pt}\slash+\omega}{k^2+\om^2-i\epsilon}\overline{\psinorm}(r^\prime )\nonumber\\
\ee
Note that only in this section, we have added an
 extra factor of $i$ in the gamma matrix in comparison to (\ref{GF11F22}) and replaced $\om$ by $-\om$
 to make the comparison with standard results easier.
Inserting (\ref{X5X})  in (\ref{SIM}), we obtain the on-shell one-loop effective action
${\cal S}_F[A_{\mu}^{(0)}\equiv n_{\mu}]=n_{\mu}n_{\nu}{\rm Im}\,\tilde{G}_{F}^{\mu\nu}( q)$,

\begin{widetext}
\be
\label{SIMX}
n_{\mu}n_{\nu}{\rm Im}\,\tilde{G}_{F}^{\mu\nu}( q)&\approx &\int \frac{d^3k}{(2\pi)^3}\int _{-|q_0|}^0\frac {dk^0}{2\pi}\int dr\sqrt{-g}\nonumber\\
&&\times \,\left(n_\mu n_\nu \frac{q^4R^6}{r^4} [K_0^2(qR^2/r) + K_1^2(qR^2/r)]+{q_\mu q_\nu}n^2 \frac{q^2R^6}{r^4} K_1^2(qR^2/r)\right)
\nonumber\\
&&\times {\rm Tr}\left(\,{\rm Im}\,D_R(r,r, k)\Gamma^{\mu}(-ik)_\alpha g^{\alpha\nu}\right)\,{\rm Im}{\cal V}_R|_{t=0}\nonumber\\
&\approx &C_{\lambda}\,\int dr\sqrt{-g}\sqrt{g^{ii}}g^{ii}\nonumber\\
&&\times \int_{0}^{q^2}\frac{d\om^2}{2}\int \frac{d^3k}{(2\pi)^3}\frac{n_F(\omega, \vec k)}{2E_{k}}
\left((n\cdot k)^2\frac{q^4R^6}{r^4} [K_0^2(qR^2/r) + K_1^2(qR^2/r)]+\frac{q^4}{4x^2}n^2 \frac{q^2R^6}{r^4} K_1^2(qR^2/r)\right)\nonumber\\
&&\times\,(\om R^2/r)^{2mR-1}(R^2/r)^{5}\,\sum_{j=1}^\infty\,\frac{2\sqrt{4\pi\lambda}}{s}(r/R^2)^3\delta(r-r_j)\nonumber\\
&\approx & C_{\lambda}\left(\frac{1}{q^2}\right)^{\tau-1}\int_{0}^{q^2}\frac{d\om^2}{2}(\om^2)^{\tau-2}\int \frac{d^3k}{(2\pi)^3}\frac{n_F(\omega, \vec k)}{2E_{k}}\sum_{j=1}^\infty \frac{w_j}{2j}\,w_j^{2\tau+3}\nonumber\\
&&\times \left(\frac{(n\cdot k)^2}{q^2} [K_0^2(w_j) + K_1^2(w_j)]+\frac{1}{4x_k^2}n^2 K_1^2(w_j)\right)\nonumber\\
&\approx &C_{\lambda}\left(\frac{1}{q^2}\right)^{\tau-1}\int_{0}^{q^2}\frac{d\om^2}{2}(\om^2)^{\tau-2}\int \frac{d^3k}{(2\pi)^3}\frac{n_F(\omega, \vec k)}{2E_{k}}\int_{0}^{\infty} dw\,w^{2\tau+3}\nonumber\\
&&\times\left(\frac{(n\cdot k)^2}{q^2} [K_0^2(w) + K_1^2(w)]+\frac{1}{4x_k^2}n^2 K_1^2(w)\right)\nonumber\\
&\approx &C_{\lambda}\left(\frac{1}{q^2}\right)^{\tau-1}\int_{0}^{q^2}\frac{d\om^2}{2}(\om^2)^{\tau-2}\int \frac{d^3k}{(2\pi)^3}\frac{n_F(\omega, \vec k)}{2E_{k}}\left(\frac{(n\cdot k)^2}{q^2} [I_{0,2\tau+3} + I_{1,2\tau+3}]+\frac{1}{4x_k^2}n^2 I_{1,2\tau+3}\right)\nonumber\\
\ee
\end{widetext}
Here  $E_k=({|\vec k|^2+\om^2})^{\frac 12}<|q^0|$, $k^0=E_{k}$, and $x_k=-\frac{q^2}{2k\cdot q}$.
Also we have set $r_j=R\sqrt{\alpha's}/2\sqrt{j}$, $w_j=qR^2/r_j=qz_j$,

\be
C_\lambda=\frac{\pi c_{5}R^4}{2\sqrt{4\pi\lambda}} \frac{2^{-(2mR-1)}}{\Gamma^2(mR+1/2)}
\ee
and defined the integrals

\be
I_{j,n}= \int_0^\infty dw\,w^n K_j^2(w) = 2^{n-2}
\frac{\Gamma(\nu + j)\Gamma(\nu - j)\Gamma(\nu)^2}{\Gamma(2\nu)}\nonumber\\
\ee
with $\nu = \frac{1}{2}(n+1)$. They are related to each other recursively  $(n-1)I_{1,n}=(n+1)I_{0,n}$.

The structure functions of the nuclei at small-x in the probe limit  are given by

\begin{widetext}
 \be \label{GLX1}
\tilde{F}_{2}(x_A,q^2)&&\approx 2\pi C_{\lambda}\left(\frac{1}{q^2}\right)^{\tau-1}\,\frac{1}{2x_A\,(n\cdot P_{A})^2}\int_{0}^{\Lambda^2<q^2}\frac{d\om^2}{2}(\om^2)^{\tau-2}\int \frac{d^3k}{(2\pi)^3}\frac{n_F(\omega, \vec k)}{2E_{k}}(n\cdot k)^2[I_{0,2\tau+3} + I_{1,2\tau+3}]\, \nonumber\\
\tilde{F}_{1}(x_A,q^2)&&\approx  2\pi C_{\lambda}\left(\frac{1}{q^2}\right)^{\tau-1}\int_{0}^{\Lambda^2<q^2}\frac{d\om^2}{2}(\om^2)^{\tau-2}\int \frac{d^3k}{(2\pi)^3}\frac{n_F(\omega, \vec k)}{2E_{k}}\left(\frac{1}{4x_k^2}I_{1,2\tau+3}\right)\,.
\ee
\end{widetext}
The effects caused by the diffusion in the radial direction on the structure functions far from the black-hole at low-x,
are discussed in Appendix E.

\subsection{Normalized structure functions}

To normalize the structure functions in the probe limit, we recall that  the bulk density and bulk energy density
follows from the holographic principle as

\be
\tilde{n}(qz\ll 1)=&&\int_0^{\Lambda^2}\frac{d\om^2}{2}I_K(qz\ll 1,\om)\int_0^{k_F(\om)}\frac{d^3k}{(2\pi)^3}\nonumber\\
\tilde{\epsilon}(qz\ll 1)=&&\int_0^{\Lambda^2}\frac{d\om^2}{2}I_K(qz\ll 1,\om)\nonumber\\
&&\times \int_0^{k_F(\om)}\frac{d^3k}{(2\pi)^3}\sqrt{k^2+\om^2}
\label{N1}
\ee
where $k_{F}(\om)=\sqrt{\mu_e^2-\om^2}$, and

\be
\label{IKZQ}
I_K(qz\ll 1,\om)=&&R^4\int_0^{z_{-}}dz\,z^{2\tau-3}(\om^2)^{\tau-2}\nonumber\\
=&&\frac{1}{2}\frac{R^4}{\tau-1}(\om^2 z_{-}^2)^{\tau-1}\om^{-2}
\ee
after taking $qzK_1(qz\ll 1)\approx 1$. The $\omega$-integration in (\ref{N1}) is carried over the bulk spectral density
$I_K(0,\omega)$ with an  upper  cut-off $\Lambda$. In the conformal case, the cutoff is a priori arbitrary. In the conformally
broken case, say a hard wall at $z=z_-$, we can set $z_-\Lambda=z_-m_N$ in (\ref{N1}) and assuming $\Lambda$ large,
to pick only the nucleon ground state.
A higher cutoff would include higher excited states of the nucleon.
With this in mind, we can first undo the k-integration by approximating it near the Fermi surface, and then undo the
$\omega$-integration by keeping only the leading  contribution for $\frac 1\beta\equiv (z_-m_N)^2>1$,

\be
\label{NEX}
\tilde{n}(qz\ll 1)\approx && \frac{1}{8\pi^2}\frac{R^4}{(\tau-1)\beta^{\tau-1}}\,k_F^{3}\nonumber\\
\tilde{\epsilon}(qz\ll 1)\approx && \frac{1}{8\pi^2}\frac{R^4}{(\tau-1)\beta^{\tau-1}}\,k_F^{3}E_{F}\nonumber\\
\ee
with $E_F=(k_F^2+m_N^2)^{\frac 12}$.  We now identify the bulk density $\tilde n=A/V_A$ as  the density
of a fixed target say a  nucleus, with A-nucleons in a fixed volume $V_A$ and
 a total energy  $E_A=AE_F$. The normalized structure $F_{1,2}$ are then related to our
 earlier and un-normalixed structure functions $\tilde{F}_{1,2}$ through

\be
\label{F12X}
F_{1,2}\equiv 2E_AV_A {\tilde{F}}_{1,2}  = 2AE_A \frac{{\tilde{F}}_{1,2}}{\tilde n}
\ee
\\
\\
{\bf Case-1 (Large-x): }
\\
\\
The normalized structure functions at large-x follow by inserting  (\ref{GLX1}) and (\ref{NEX}) into (\ref{F12X})
with $\beta=1/(m_Nz_-)^2$

\begin{widetext}
 \be
\frac{F_{2}(x,q^2)}{A}&&\approx 8\pi^2(\tau-1)^2e_R^2\Big(\frac{\beta m_N^2}{q^2}\Big)^{\tau-1}
\,x_{{F}}^{\tau+1}(1-x_{{F}})^{\tau-2}\,\nonumber\\
\frac{F_{1}(x,q^2)}{A}&&\approx 4\pi^2(\tau-1)^2e_R^2\Big(\frac{\beta m_N^2}{q^2}\Big)^{\tau-1}\,A
\,x_{{F}}^{\tau+1}(1-x_{{F}})^{\tau-2}\,. \nonumber\\
 \label{GRxwt773}
\ee
\end{widetext}
We define the x-fractions

\be
x_F=&&\frac{q^2}{-2p_F\cdot q}\approx \frac{q^2}{2E_F\omega}=\frac{xm_N}{E_F}\nonumber\\
x_A=&&\frac{q^2}{-2P_A\cdot q}\approx \frac{-q^2}{2E_A\omega}=\frac {x_F}A
\ee
and note that the large-x structure functions in (\ref{GRxwt773}) in the probe approximation obey the
analogue of the Callan-Gross relation $F_2=2x_AF_1$ for a holographic and dilute nucleus.  Also we
note that (\ref{GRxwt773}) are analogous to the so-called structure functions of the nucleus obtained
through the so-called x-scaling of the structure functions of the nucleon.
\\
\\
{\bf Case-2 (Small-x): }
\\
\\
Doing the momentum integrals in~(\ref{GLX1}) near $k\rightarrow k_{F}$ and doing the appropriate normalization as in the large-x regime, we find
\begin{widetext}
 \be
\frac{F_{2}(x,q^2)}{A}&&\approx {\pi C_{\lambda}}
\Big(\frac{\beta m_N^2}{q^2}\Big)^{\tau-1}\,\frac{1}{x_{{F}}}[I_{0,2\tau+3} + I_{1,2\tau+3}]\,\nonumber\\
\frac{F_{1}(x,q^2)}{A}&&\approx {\pi C_{\lambda}}\Big(\frac{\beta m_N^2}{q^2}\Big)^{\tau-1}\,A
\,\left(\frac{1}{4x_F^2}I_{1,2\tau+3}\right)\,, \nonumber\\
 \label{GLX2}
\ee
\end{widetext}

\subsection{R-ratio in the probe limit}

We define the R-ratio of the nucleus in the probe (dilute) limit as

\be
\label{rr-dilute}
R_{\rm dilute}(x,q^2)\equiv\frac{\frac 1A F_2^{\rm dilute}(x,q^2)}{F_2^{\rm nucleon}(x,q^2)}
\ee
where $F_2^{\rm dilute}(x,q^2)$ is given by the sum of (\ref{GRxwt773}) for large-x and (\ref{GLX2}) for small-x,

\begin{widetext}
\be
\frac{F_{2}^{\rm dilute}(x,q^2)}{A}\approx f\Big(\frac{\beta m_N^2}{q^2}\Big)\times\Big(\frac{\beta m_N^2}{q^2}\Big)^{\tau-1}\Big(
8\pi^2
(\tau-1)^2 e_R^2\,x_{{F}}^{\tau+1}(1-x_{{F}})^{\tau-2}
+\frac{c_{5}\pi^2 }{2\sqrt{4\pi\lambda}} \, \frac{[I_{0,2\tau+3} + I_{1,2\tau+3}]}{2^{(2\tau-4)}\Gamma^2(\tau-1)}
\,\frac{1}{x^j_{{F}}} \Big)\,,\nonumber\\
 \label{f2dilute}
\ee
\end{widetext}
and $F_2^{\rm nucleon}(x,q^2)$ is given by

\begin{widetext}
\be
\frac{F_{2}^{\rm nucleon}(x,q^2)}{A}=f\Big(\frac{\beta m_N^2}{q^2}\Big)\times\Big(\frac{\beta m_N^2}{q^2}\Big)^{\tau-1}\Big(
8\pi^2
(\tau-1)^2 e_R^2\,x^{\tau+1}(1-x)^{\tau-2}
+\frac{c_{5}\pi^2 }{2\sqrt{4\pi\lambda}} \, \frac{[I_{0,2\tau+3} + I_{1,2\tau+3}]}{2^{(2\tau-4)}\Gamma^2(\tau-1)}
\,\frac{1}{x^j} \Big)\,,\nonumber\\
 \label{f2nucleon}
\ee
\end{widetext}
The additional multiplicative function $f(\iota )$ with argument $\iota ={\beta m_N^2}/{q^2}$ asymptotes a  constant 
 in the  DIS regime which is our AdS$_5$ result. The specific form of $f(\iota )$ depends on the details of the confining model for finite
$q^2$ (see for example Eq.~3.28 and Eq.~4.3 in~\cite{Braga:2011wa} for the soft-wall model) but in the  DIS limit, the asymptotic constant
can be reabsorbed through a shift $\beta\rightarrow \tilde\beta$.
We recall that $\beta m_N^2=1/z_-^2$ is related to the confining scale here, and that $x_FE_F=xm_N$. Also we have 
in (\ref{GLX2}) $j=1$ in the absence of transverse diffusion or curvature corrections. When the latters are included
$j\rightarrow 1-{\cal O}(1/\sqrt\lambda)$.  The structure function of the proton follows
from (\ref{f2dilute})  by setting $k_F=0$ or through the substitution $x_F\rightarrow x$. The R-ratio for the probe or dilute limit
is independent of $\beta\rightarrow \tilde\beta$ and $q^2$. Note that the first contribution is proportional to $e_R^2\lambda^0$  while the second is proportional
to $e_R^0/\sqrt{\lambda}$ independent of the R-charge, as expected for Pomeron-like exchange.

\begin{figure}[!htb]
 \includegraphics[height=5cm]{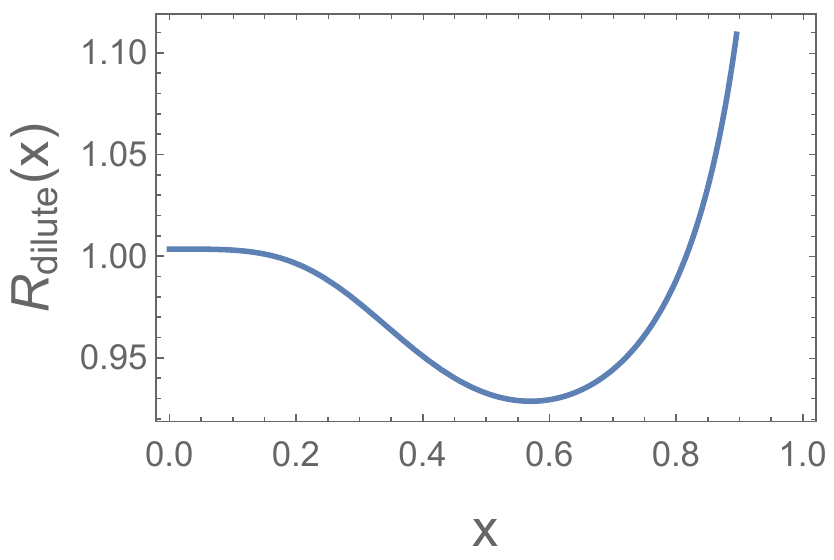}
  \caption{Dilute R-ratio (\ref{rr-dilute}) for $k_F/m_N=0.3$. See text.}
\label{dilute-ratio}
\end{figure}

In Fig.~\ref{dilute-ratio} we show the behavior of the dilute R-ratio (\ref{rr-dilute}) versus $x$ for fixed Fermi momentum $k_F/m_N=0.3$.
The holographic parameters used are fewer but consistent with those used for the dense R-ratio in Fig.~\ref{dense-ratio1}. Specifically, we have used:
$e_R=0.3$ (R-charge of the bulk fermion), $2\pi^2c_5/\sqrt{4\pi\lambda}=0.01$ (strong coupling), $\tau=3$ (hard scaling exponent),
$j=0.08$ (Pomeron intercept). In the dilute case, the R-ratio is dominant at large-x and asymptotes 1 at small-x. Clearly visible is the
EMC-like effect for $0.2<x<0.8$ and the Fermi motion for $x>0.8$. We have checked that the overall features of Fig.~\ref{dilute-ratio}
remain unchanged for smaller values of $e_R$ but fixed $k_F/m_N$ in conformity with the probe limit. This holographic behavior is
very similar to the one we presented  recently using general arguments~\cite{Mamo:2018eoy}.

\section{Hybrid model for DIS scattering on a finite nucleus}

Nuclei to a large extent are a collection of nucleons trapped by a mean-field usually the result of a Hartree-Fock approximation to
the two-body  and dominant interaction. Three- and higher-body interactions are suppressed as expected from the saturation
properties and the bulk compressibility of nuclear matter.

DIS scattering on a nucleus is expected to be dominated by incoherent
scattering  on a dilute collection of nucleons at intermediate- and large-x, with modifications at small-x due coherent
scattering induced by the residual and small two- and three-body mediated interactions following the rapid growth in parton-x.

 A way to capture
this, short of a more systematic density expansion outlined in~\cite{Mamo:2018eoy}, is to suggest that the nucleus structure function $F_2^{nucleus}(x,q^2)$ is composed of
the dilute contribution (\ref{f2dilute}) warped by AdS$_5$  (low density and dominant contribution) plus the dense and quantum corrected black-hole
(\ref{f2dense}) (high density and subdominant contribution) with a new and arbitrary mixing coefficient $\mathbb C_{2}$ to be determined from the best  fit to the
world-data, i.e.,

\be\label{f2nucleus}
\frac{F_2^{nucleus}(x,q^2)}{A}=\frac{F_2^{dense}(x,q^2)}{A}+ \mathbb C_{2} \frac{F_2^{dilute}(x,q^2)}{A}\,.\nonumber\\
\ee
The nucleus  R-ratio is then

\be \label{rnucleus}
R_{nucleus}(x,q^2)\equiv\frac{\frac{1}{A}F_2^{nucleus}(x,q^2)}{F_2^{nucleon}(x,q^2)}\,.
\ee

To compare the holographic results following from (\ref{rnucleus}) to the world-data from DIS scattering on heavy and light nuclei,
all the holographic parameters are set as before:
 $\tilde\alpha=N_c/4N_f=1$ (ratio of branes), $2\pi^2c_5/\sqrt{4\pi\lambda}=0.01$
(strong coupling) and $e_R=0.3$ (charge of the probe fermions),  $\tau=3$ (hard scaling law), $j=0.08$ (Pomeron intercept) and
$\beta\rightarrow \tilde\beta=17.65$ (confining scale). The parameters of the emergent Fermi surface are also fixed as before: $v_F=1$ (Fermi velocity),  $\tilde{h}_2=1$ for simplicity,
$\mu/m_N=1.2$ (chemical potential  for a typical nucleus), and $k_F/m_N=0.395$ (for Au, Pb, and Fe), 0.277 (for C), 0.119 (for He) (Fermi momentum).  We note that our choices for the nuclei Fermi momenta are very close to those extracted from quasielastic electron scattering experiments on nuclei~\cite{Moniz:1971mt} modeled using the Fermi gas model~\cite{Moniz:1969sr}.  The parameter $\mathbb C_{1}=0.07$ fixes the quantum correction to the black hole due to the far horizon contribution. Our analysis of
the data shows that the mixing parameter for the best fit is $\mathbb C_{2}=0.793$. Clearly, there is some form of double counting of the quantum corrections at large-x, but this
is minimal since $\mathbb C_1/\mathbb C_2=0.09$.

We show in Figs.~\ref{dense-ratio3-gold-lead}-\ref{dense-ratio3-helium3}a, the  nucleus R-ratio versus x for Au, Pb, Fe, C, and He
on a linear scale. The overall agreement with the heavy and heavy-light nuclei data is fair throughout, the exception being He where the holographic model 
overshoots in the anti-shadowing region. This maybe an indication  that the coherent scattering captured by the black-hole should be weaker, which is sensible.
Figs.~\ref{dense-ratio3-gold-lead}-\ref{dense-ratio3-helium3}b display the same ratio on a semi-logarithmic scale to highlight the low-x contribution where the data are  scarce.  A key proposal of the future Electron-Ion-Collider (EIC) is to provide measurements for the nuclear R-ratio in this region.

Remarkably, the R-ratio (\ref{rnucleus}) exhibits shadowing for $x<0.1$, anti-shadowing for $0.1<x<0.3$, the EMC-like effect for $0.3<x<0.8$ and Fermi motion for $x>0.8$.  As we noted earlier, shadowing and anti-shadowing are  caused by coherent DIS scattering on the dense component in (\ref{f2nucleus}) due to the underlying black-hole  in our analysis, while the EMC effect and Fermi motion are  mostly due to incoherent DIS scattering on the  dilute component  in (\ref{f2nucleus}) with the nucleus mostly composed 
of individual nucleons warped by AdS$_5$.

\begin{widetext}

\begin{figure}[!htb]
\begin{center}
\begin{tabular}{cc}
\includegraphics[height=5cm]{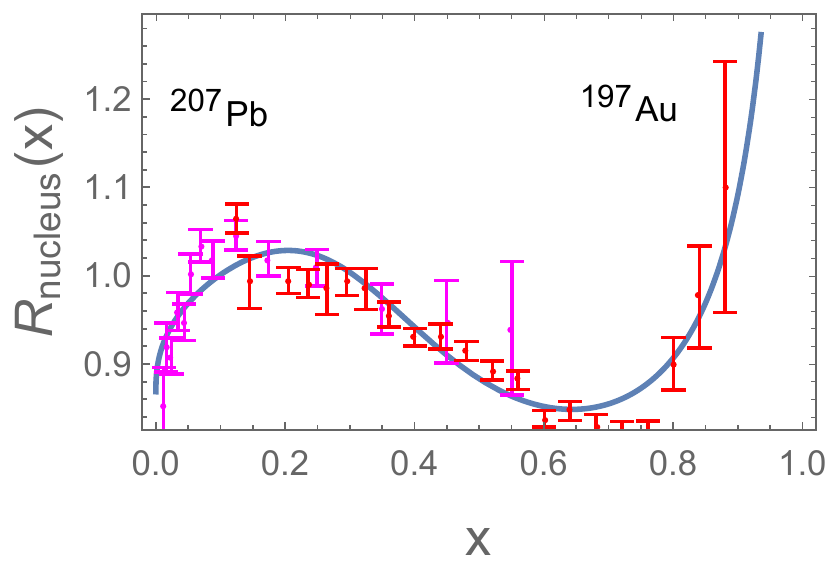}
&
\includegraphics[height=5cm]{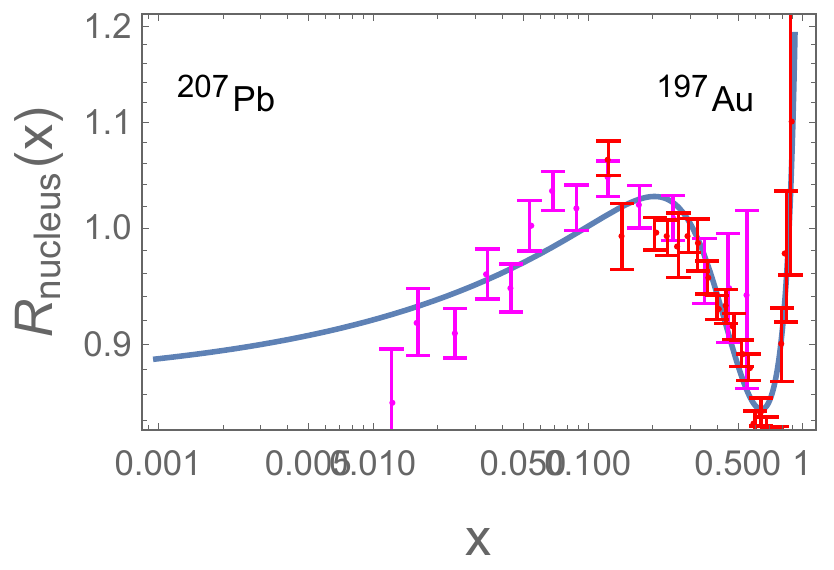}

\\
(a) & (b)
\end{tabular}
\caption{Nucleus R-ratio (\ref{rnucleus}) for Lead Pb with A=207 and Gold Au with A=197. We have fixed $k_F/m_N=0.395$. See text. The data points are from \cite{Amaudruz:1995tq, Arneodo:1996ru} for Pb (pink), and \cite{Gomez:1993ri} for Au (red), see also Fig.7 in \cite{Malace:2014uea}.}
\label{dense-ratio3-gold-lead}
\end{center}
\end{figure}

\begin{figure}[!htb]
\begin{center}
\begin{tabular}{cc}
\includegraphics[height=5cm]{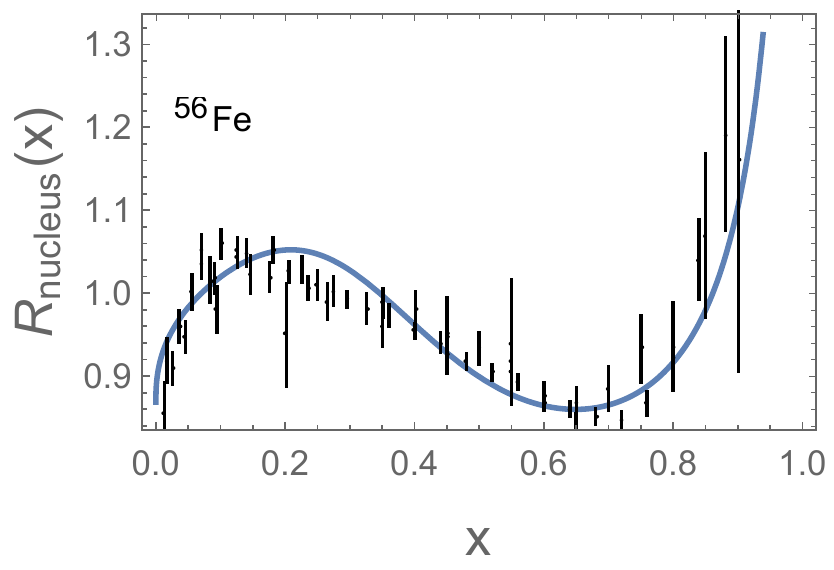}
&
\includegraphics[height=5cm]{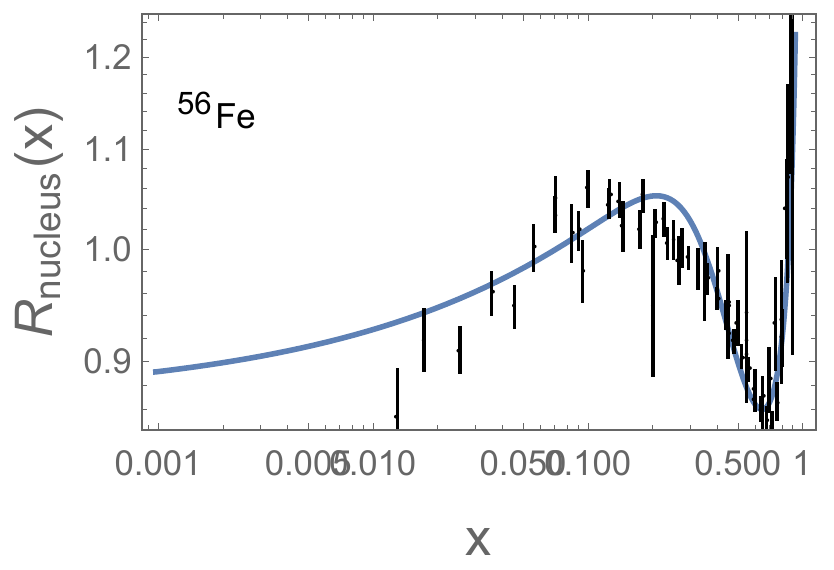}

\\
(a) & (b)
\end{tabular}
\caption{Nucleus  R-ratio (\ref{rnucleus}) for Iron Fe with A=56. We have fixed $k_F/m_N=0.395$. See text. The data points are from \cite{Amaudruz:1995tq, Arneodo:1996ru, Bodek:1983qn, Benvenuti:1987az}, see also Fig.5 in \cite{Malace:2014uea}.}
\label{dense-ratio3-iron}
\end{center}
\end{figure}

\begin{figure}[!htb]
\begin{center}
\begin{tabular}{cc}
\includegraphics[height=5cm]{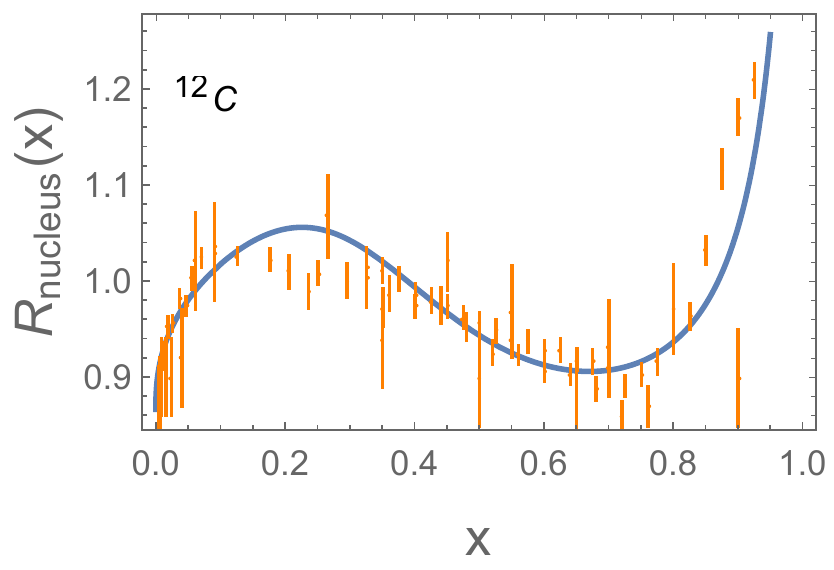}
&
\includegraphics[height=5cm]{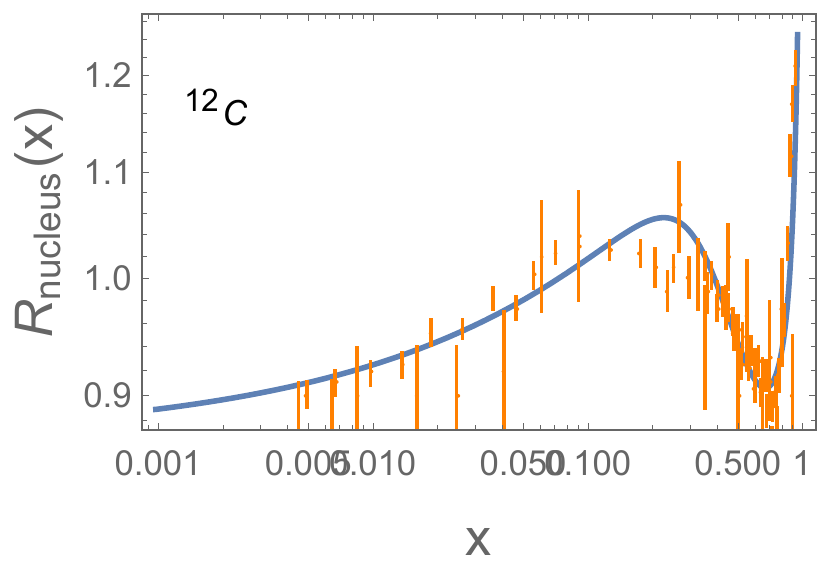}

\\
(a) & (b)
\end{tabular}
\caption{Nucleus R-ratio (\ref{rnucleus}) for Carbon C with A=12. We have fixed $k_F/m_N=0.277$.  See text. The data points are from \cite{Gomez:1993ri, Seely:2009gt, Arneodo:1995cs, Arneodo:1989sy}, see also Fig.3 in \cite{Malace:2014uea}.}
\label{dense-ratio3-carbon}
\end{center}
\end{figure}

\begin{figure}[!htb]
\begin{center}
\begin{tabular}{cc}
\includegraphics[height=5cm]{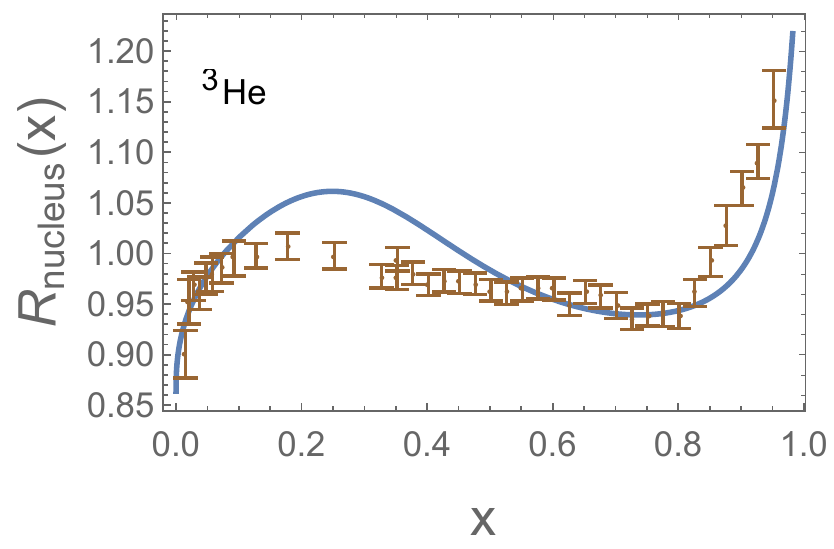}
&
\includegraphics[height=5cm]{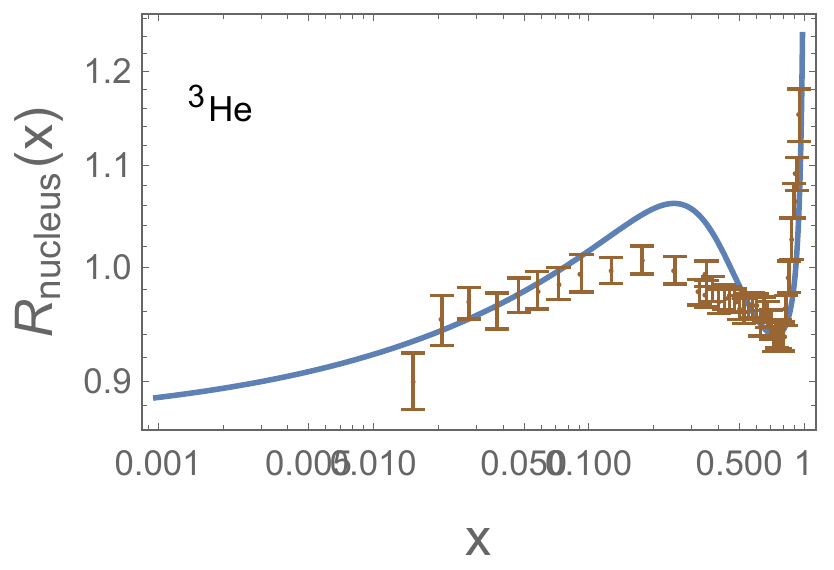}

\\
(a) & (b)
\end{tabular}
\caption{Nucleus R-ratio (\ref{rnucleus}) for Helium He with A=3. We have fixed $k_F/m_N=0.119$. See text. The data points are from \cite{Seely:2009gt, Ackerstaff:1999ac}, see also Fig.1 in \cite{Malace:2014uea}.}
\label{dense-ratio3-helium3}
\end{center}
\end{figure}

\end{widetext}

\section{Conclusions}

In the double limit of a large number of colors and strong coupling, DIS scattering off
an extremal black hole is of order $N_c^2$ following from the absorption of the bulk
R-current by the black hole. The corresponding 
structure functions are dominated by low-x. Scattering off the black hole is the ultimate
coherent scattering off a dense nucleus with strong shadowing as we noted in~\cite{Mamo:2018ync}.

To order $N_c^0$, DIS scattering is off holographic fermions hovering near the
black-hole horizon due to quantum pair creation, and warped holographic fermions
far from the black hole horizon near the boundary. Close to the horizon, the geometry is that of
AdS$_2\times$R$^3$ with an emergent Fermi surface and anomalous scaling laws.
DIS scattering off these bulk fermions show that their  structure functions on
the boundary exhibit anomalous exponents and modified hard scattering rules in
comparison to scattering off bulk fermions  in the dilute or probe limit. DIS scattering
off  these fermions exhibit Fermi motion at large-x.


 DIS scattering on a wide range of nuclei maybe captured by a hybrid holographic model
whereby most of the DIS scattering is incoherent and off a holographic Fermi liquid warped 
by AdS$_5$  at intermediate- and large-x, with the remainder following from coherent scattering 
off  a quantum corrected RN-AdS black hole at small-x. The ensuing results
agree remarkably with the existing data on DIS scattering on finite  nuclei over a broad range
of parton-x. More data at low-x from the planned EIC collider will be welcome for a better understanding of the coherent scattering through the black-hole mechanism suggested here.

\section{Acknowledgements}
This work was supported by the U.S. Department of Energy under Contract No.
DE-FG-88ER40388.

\appendix

\section{Conventions in curved space}

The gamma matrices in curved and tangent space used to analyze the Dirac equation
in the extremal RN-AdS black hole will be made explicit here. For that, consider the generic line element
in curved space

\be
\label{metric}
 ds^2 = -g_{tt} dt^2 + g_{rr} dr^2 + g_{ii} dx_i^2
 \ee
If we refer to the indices in curved space by $\mu, \nu$ (also $t,i$) and those in the tangent space by $a,b$
(also $\underline t, \underline i$) then the gamma matrices are related by

 \be
 \Ga^\mu = \Ga^a e_a{^\mu} , \qquad \Ga^r = \sqrt{g^{rr}}\, \Ga^{\ur} \ .
 \ee
If we set the vierbeins as~\cite{Faulkner:2013bna}

\be
 e^\ut =
  \sqrt{g_{tt}}\, dt ,\quad
  e^\ui = \sqrt{g_{ii}}\, dx^i \,,\nonumber\\
\ee
then we have

\be
 \Ga^t =\sqrt{g^{tt}}\, \Ga^{\underline{t}},
 \qquad
 \Ga^i=\sqrt{g^{ii}} \,\Ga^{\underline{i}},
 \qquad
 \Ga^r = \sqrt{g^{rr}}\, \Ga^{\ur} \ .\nonumber\\
 \ee
In the tangent space, the gamma matrices read~ \cite{Gubser:2012yb}

\be
&&\Gamma^{\underline{t}}=\begin{pmatrix}
i\sigma_1 & 0\\
0 & i\sigma_1
\end{pmatrix}\qquad
\Gamma^{\underline{r}}=\begin{pmatrix}
\sigma_3 & 0\\
0 & \sigma_3
\end{pmatrix}\nonumber\\
&&\Gamma^{\underline{x}}=\begin{pmatrix}
-\sigma_2 & 0\\
0 & \sigma_2
\end{pmatrix}\qquad
\Gamma^{\underline{y}}=\begin{pmatrix}
0 & -\sigma_2\\
-\sigma_2 & 0
\end{pmatrix}\nonumber\\
&&\Gamma^{\underline{z}}=\begin{pmatrix}
0 & i\sigma_2\\
-i\sigma_2 & 0
\end{pmatrix}.
\ee
The non-vanishing spin connections are

 \bea
\om_{\ut \ur} = -f_0 e^\ut\,,
 \qquad
 \om_{\ui \ur } = f_1 e^{\ui}
 \eea
with
\be
f_0 \equiv \ha {g_{tt}' \ov \tilde g_{tt}} \sqrt{g^{rr}}\,, \qquad
f_1 \equiv \ha{g_{ii}' \ov g_{ii}} \sqrt{g^{rr}}\,.
\ee

\section{Soft spinors}


The soft normalizable wavefunctions were constructed in~\cite{Faulkner:2013bna}, we reproduce them here
for completeness. The Dirac equation in the AdS$_2\times$R$^3$ geometry is solved by the rescaled spinors

\be
\label{swf}
&&\psinorm_{1} (r;\vk)=(- g g^{rr})^{-{1 \ov 4}}\left(\begin{matrix}\Phinorm_{1}\\0\\0\end{matrix}\right)\times \sqrt{\frac{r_{-}}{R^2}}\times\Big(\frac{r_{-}}{R^2}\Big)^{\nu_k}\,,\nonumber\\
&&\psinorm_{2} (r;\vk)=(- g g^{rr})^{-{1 \ov 4}}\left(\begin{matrix}0\\0\\\Phinorm_{2}\end{matrix}\right)\times\sqrt{\frac{r_{-}}{R^2}}\times\Big(\frac{r_{-}}{R^2}\Big)^{\nu_k}\,\nonumber\\
\ee
with

\be
\label{sout}
 &&\Phinorm_{1} =  \left(\begin{matrix}v_1\\v_2\end{matrix}\right)\equiv {1 \ov W} a_+(k_0,k) \Psi^{(0)}_-\,\nonumber\\
&& \Phinorm_{2} = \left(\begin{matrix}\tilde{v}_1\\\tilde{v}_2\end{matrix}\right)\equiv  \Phinorm_{1}(\vk\mapsto -\vk)
\ee
and $W = - i v_+^T \sig^2 v_-$, $a_+ (k_0,k) = \tilde{c}_1 (k-k_F) + \tilde{c}_2 k_0 + \cdots$, where $\tilde{c}_{1,2}\sim \frac{R^2}{r_{-}}$.
The  explicit spinors are

\be
\Psi_{\pm }^{(0)} &\approx &  v_\mp   \le({ r-r_-\ov R_2^2} \ri)^{\pm\nu_k} \nonumber\\
v_\pm &=&   \left( \begin{matrix} m R_2 \pm \nu_k
\cr {k R \ov r_-} R_2 + e_R \sqrt{\frac{\tilde{\alpha}}{3}}
\end{matrix} \right) \,,
\ee
with $R_2=R/2\sqrt{3}$, and
\bea
\nu_k &=& \sqrt{m_k^2 R_2^2 - \frac{\tilde{\alpha}}{3} e_R^2} , \quad m_k^2 \equiv m^2 +
{k^2 R^2 \ov r_-^2}\ .\nonumber\\
 \eea
Note that for pure AdS$_2$, the soft wave-functions simplify

\be
\label{sout-pure}
 \Phinorm_{1} = \Psi^{(0)}_-\,\qquad
 \Phinorm_{2} =  \Phinorm_{1}(\vk\mapsto -\vk)\,.
\ee

Finally, note that the Feynman propagator for the soft part in (\ref{spinorG20}-\ref{spinorG2}) is given by

\be
D_{F}(r,r^\prime ;k) = \psinorm_{\alpha}(r, k) \,\left(\int \frac{d{\omega}}{2\pi} \frac{\rho_B^{\alpha\gamma}(\omega,\vec{k}) }{k^0 - \omega}\right)\,
\overline{\psinorm_{\gamma}} (r^\prime,  k)\nonumber\\
\label{spectrep}
\ee
with the boundary spectral function $\rho_B(\omega,\vec{k})$, and the normalizable wave function $\psinorm_{\alpha}(r, \vec k)$ for the Dirac equation in curved AdS$_5$.

\section{Parameters $v_F,h_1,h_2,k_F/\mu$ entering  the  near horizon Green's function}

We recall that the  retarded Green's function (\ref{spinorG}) in the near horizon limit can be recast in the form

\be
 \mathcal{G}^{11}_{R}(k^0,\vk) \approx && {\tilde{h}_1\mu \ov  k-\big(\frac{k_F}{\mu}\big)\mu - \frac 1{v_F}k^0-\Pi(k^0)}\begin{pmatrix}
0 & 0\\
0 & 1
\end{pmatrix}\nonumber\\
\Pi(k^0)=&&\tilde{h}_2\mu\, e^{i \ga_{k_F}}\, \Big(\frac{k^0}{\mu}\Big)^{2 \nu_{k}}\,,\nonumber\\
\nu_k=&&\sqrt{\tilde{\alpha}}\left(\frac{k^2}{\mu^2}-\frac{k_R^2}{\mu^2}\right)^{\frac 12}\,,\nonumber\\
\ga_k = && \arg\left( \Gamma(-2\nu_k) \le(e^{-2 \pi i \nu_k} - e^{- \frac{2 \pi}{\sqrt{3}} e_R \sqrt{\tilde{\alpha}}} \ri)\right)\,,\nonumber\\
 \ee
where ${k_R^2}/{\mu^2}=\left(e_R^2\tilde{\alpha}-\frac{1}{4}(\tau-\frac{3}{2})^2\right)/{3\tilde{\alpha}}$,
and ${k_0}/{\mu}\sim {C_0}/{2\sqrt{3}\sqrt{\tilde{\alpha}}}$ with the dimensionless constant

\be
C_0(\nu_{k_F})=(v_F \tilde{h}_2{\rm Im}\,(e^{i\gamma_{k_F}}(-1)^{2\nu_{k_F}}))^{\frac{1}{1-2\nu_{k_F}}}\nonumber\\
\ee

The dimensionless parameters $v_F$, $\tilde{h}_1$, $\tilde{h}_2$, and ${k_F}/{\mu}$ that characterize the
retarted Green's function (\ref{spinorG}) and (\ref{spinorG2}), are in principle  determined numerically as in~\cite{Faulkner:2009wj} for the 3-dimensional spacetime, and analytically for R-charged black holes as in~\cite{Gubser:2012yb}. Here, we will not carry out the numerical analysis to determine the parameters precisely,
 but we note that all of them are some functions of $e_R$, $\tau$, and $\tilde{\alpha}$, i.e.,
 $v_F(e_R,\tau,\tilde{\alpha})$, $\tilde{h}_1(e_R,\tau,\tilde{\alpha})$,  $\tilde{h}_2(e_R,\tau,\tilde{\alpha})$, and ${k_F}/{\mu}(e_R,\tau,\tilde{\alpha})$. Therefore, the dimensionless parameters $v_F,\tilde{h}_1$, $\tilde{h}_2$, ${k_F}/{\mu}$, and the dimensionful parameter $\mu$ are free parameters. The parameter
 $\tilde{h}_1$ drops out in the normalization.  The remaining parameter is set to $\tilde{h}_2=1$ for simplicity.

\section{Effective vertices}

The soft-to-hard transition vertices entering in the bulk DIS amplitude  involve
(\ref{sout-pure}) for the reduction to AdS$_2$ or  (\ref{sout}) in general for the soft part,
with the hard part of the wave-function given by

\be
&&u_1=\Big(\frac{R^2}{r}\Big)^\frac{5}{2}J_{mR-\frac{1}{2}}\Big(\omega_1\frac{R^2}{r}\Big)\nonumber\\
&&u_2\equiv 0
\ee
More specifically, for pure AdS$_2$, the transition vertex is simply given by

\bea \label{vertex-pure}
 &&\Lambda_{11}^x (z_{2};\om_1;q;k)=\nonumber\\
 && C(\nu_{k})\,\Big(\frac{r_{-}}{R^2}\Big)^{\nu_k+\frac{1}{2}}\,q\int_{0}^{\infty} dz_{2}z_{2}^{\frac{3}{2}+\nu_{k}}K_{1}(qz_{2})J_{mR-\frac{1}{2}}(\om_1 z_{2})\,,\nonumber\\
\eea
with
\be \label{cvertex-pure}
C(\nu_{k})=e_RR^2(mR_{2}+\nu_{k})(2\sqrt{3})^{-\nu_{k}}\,.
\ee

In general, the transition vertices are of the form

\begin{widetext}
\bea \label{vertex211}
 \Lambda_{11}^i (r_{2};\om_1;q;k)&=&-e_R\,\Big(\frac{r_{-}}{R^2}\Big)^{\nu_k+\frac{1}{2}}\, \int dr_{2} \sqrt{-g}(- g g^{rr})^{-{1 \ov 4}}\sqrt{g^{ii}}K_{A}(r_2;q)\, u_{1}\left(\begin{matrix}0\\1\\0\\0\end{matrix}\right)^{T} \, \Gamma^{\underline{i}} \, \left(\begin{matrix}v_1\\v_2\\0\\0\end{matrix}\right)\,,\nonumber\\
\Lambda_{22}^i (r_{2};\om_1;q;k)&=&-e_R \,\Big(\frac{r_{-}}{R^2}\Big)^{\nu_k+\frac{1}{2}}\,\int dr_{2}\sqrt{-g}(- g g^{rr})^{-{1 \ov 4}}\sqrt{g^{ii}} K_{A}(r_2;q)\, u_{2}\left(\begin{matrix}0\\0\\0\\1\end{matrix}\right)^{T} \, \Gamma^{\underline{i}} \, \left(\begin{matrix}0\\0\\\tilde{v}_1\\\tilde{v}_2\end{matrix}\right)\,,\nonumber\\
\Lambda_{11}^j (r_{1};k;q;\om_1)&=&-e_R \,\Big(\frac{r_{-}}{R^2}\Big)^{\nu_k+\frac{1}{2}}\, \int dr_{1} \sqrt{-g}(- g g^{rr})^{-{1 \ov 4}}\sqrt{g^{jj}} K_{A}(r_1;q) \,u_{1}\left(\begin{matrix}v_1\\v_2\\0\\0\end{matrix}\right)^{T} \, \Gamma^{\underline{i}} \, \left(\begin{matrix}0\\1\\0\\0\end{matrix}\right) \,,\nonumber\\
\Lambda_{22}^j (r_{1};k;q;\om_1)&=&-e_R \,\Big(\frac{r_{-}}{R^2}\Big)^{\nu_k+\frac{1}{2}}\, \int dr_{1} \sqrt{-g}(- g g^{rr})^{-{1 \ov 4}}\sqrt{g^{jj}} K_{A}(r_1;q) \,u_{2} \left(\begin{matrix}0\\0\\\tilde{v}_1\\\tilde{v}_2\end{matrix}\right)^{T}  \, \Gamma^{\underline{i}} \,\left(\begin{matrix}0\\0\\0\\1\end{matrix}\right)\,.\nonumber\\
\eea
\end{widetext}
Using the gamma matrices explicitly, we can simplify the effective vertices (\ref{vertex211}). More specifically, we have

\bea
\label{vertex213}
 &&\Lambda_{11}^x (r_{2};\om_1;q;k)=ie_R\,\Big(\frac{r_{-}}{R^2}\Big)^{\nu_k+\frac{1}{2}}\,\nonumber\\
 &&\times \int dr_{2} \sqrt{-g}(- g g^{rr})^{-{1 \ov 4}}\sqrt{g^{xx}} K_{A}(r_2;q) \, u_{1}v_{1}\,,\nonumber\\
 \eea
 with the rest of the vertices following by symmetry

 \bea
\Lambda_{11}^x (r_{1};k;q;\om_1)&=&- \Lambda_{11}^x (r_{2};\om_1;q;k) \,,\nonumber\\
\Lambda_{22}^x (r_{1};k;q;\om_1)&=&  \Lambda_{22}^x (r_{2};\om_1;q;k)\equiv 0\,,\nonumber\\
\eea
and all other components vanishing. Performing the change of variable
 $r={R^2}/{z}$ and setting $z\ll z_{-}$, we can re-write the integral in (\ref{vertex213}) as

\bea
\label{vertexx}
 \Lambda_{11}^x (z_{2};\om_1;q;k)=&&C(\nu_{k})a_+(k_0,k)\,I_z(\omega_1; q;k)\nonumber\\
 =&& C(\nu_{k})a_+(k_0,k)\,\Big(\frac{r_{-}}{R^2}\Big)^{\nu_k+\frac{1}{2}}\,\nonumber\\
\times && \int_{0}^{\infty} dz_{2}z_{2}^{\frac{3}{2}+\nu_{k}}qK_{1}(qz_{2})J_{mR-\frac{1}{2}}(\om_1 z_{2})\,,\nonumber\\
\eea
with

\be
\label{vertexxB}
C(\nu_{k})=e_R R^2\frac{(mR_{2}+\nu_{k})}{W}(2\sqrt{3})^{-\nu_{k}}\,,
\ee
The integration can be carried out analytically with the result

\begin{widetext}
\be \label{vertexI2}
  I_{z}(\om_1;q;k)=&&\,\Big(\frac{r_{-}}{R^2}\Big)^{\nu_k+\frac{1}{2}}\,\int_{0}^{\infty} dz_{2}z_{2}^{\frac{3}{2}+\nu_{k}}qK_{1}(qz_{2})J_{mR-\frac{1}{2}}(\om_1 z_{2})\nonumber\\
  = && \,\Big(\frac{r_{-}}{R^2}\Big)^{\nu_k+\frac{1}{2}}\, C_{z}(\nu_{k})\frac 1{q^{(\nu_{k}+\frac{3}{2})}}\Big(\frac{\om_1}{q}\Big)^{mR-\frac{1}{2}}{}_2F_1\Big(\frac{mR+\nu_{k}+3}{2},\frac{mR+\nu_{k}+1}{2},mR+\frac{1}{2},-\frac{\om_{1}^2}{q^2}\Big)\,,\nonumber\\
\ee
\end{widetext}
with

\be
 C_{z}(\nu_k)=2^{\nu_{k}+\frac{1}{2}}\frac{\Gamma(\frac{mR+\nu_{k}+3}{2})\Gamma(\frac{mR+\nu_{k}+1}{2})}{\Gamma(mR+\frac{1}{2})}
 \ee
Note that for the  special value $\nu_{k}=\nu_{k}^*=mR$, the integrand reduces to the one in~\cite{Polchinski:2002jw}, and can be evaluated exactly as

\begin{widetext}
\be \label{vertexI3}
  I_{z}(\om_1;q;k)=&&\,\Big(\frac{r_{-}}{R^2}\Big)^{\nu_k+\frac{1}{2}}\,\int_{0}^{\infty} dz_{2}z_{2}^{\frac{3}{2}+\nu_{k}^*}K_{1}(qz_{2})J_{mR-\frac{1}{2}}(\om_1 z_{2})\nonumber\\
  =&&\,\Big(\frac{r_{-}}{R^2}\Big)^{\nu_k+\frac{1}{2}}\, C_{z}(\nu_{k}^*,q)q^{-(mR+\frac{5}{2})}\Big(\frac{\om_1}{q}\Big)^{mR-\frac{1}{2}}\Big(1+\frac{\om_{1}^2}{q^2}\Big)^{-(mR+\frac{3}{2})}
\ee
\end{widetext}
and $C_{z}(\nu_{k}^*)=2^{mR+\frac{1}{2}}\Gamma(mR+\frac{3}{2})$.

\section{Low-x structure functions with radial diffusion}

Far from the black hole and including diffusion in the radial direction, the structure functions can be written as
\be
F_{2}(x_A,q^2)&&= \int_{0}^{\infty}\frac{d\om^2}{2}\int \frac{d^3k}{(2\pi)^3}\frac{n_F(\omega, \vec k)}{2E_{k}}F_2(x_{k},q^2,\om)\, \nonumber\\
F_{1}(x_A,q^2)&&=\int_{0}^{\infty}\frac{d\om^2}{2} \int \frac{d^3k}{(2\pi)^3}\frac{n_F(\omega, \vec k)}{2E_{k}}F_1(x_k,q^2,\om)\,,\nonumber\\
\ee
where~\cite{Brower:2010wf}(see also \cite{Kovensky:2018xxa})

\begin{widetext}
\be
&&F_2(x_k,q^2,\om) = \frac{g_0^2\rho^{3/2}}{32 \pi^{5/2}} \int \frac{dz}{z} \frac{dz'}{z'} P^2_{A}(z,q^2) P_{\psi}(z',\om) (zz'q^2)\,
\frac{e^{\zeta_{k}(1-\rho)}}{\sqrt{\zeta_{k}}}e^{-\frac{\log^2\left(z/z'\right)}{\rho
\zeta_{k}}} \nonumber\\
&&2x_kF_1(x_k,q^2,\om) = \frac{g_0^2\rho^{3/2}}{32 \pi^{5/2}}
\int \frac{dz}{z} \frac{dz'}{z'} P^1_{A}(z,q^2) P_{\psi}(z',\om)
(zz'q^2)\,
\frac{e^{\zeta_{k}(1-\rho)}}{\sqrt{\zeta_{k}}}e^{-\frac{\log^2\left(z/z'\right)}{\rho
\zeta_{k}}} \, ,
\label{F1brower}
\ee
\end{widetext}
with
\be
P^2_{A}(z,q^2) &&= (q z)^2 \left(K_1^2(q z) + K_0^2(q z)\right)\nonumber\\
P^1_{A}(z,q^2) &&= (q z)^2 K_1^2(q z)\nonumber\\
P_{\psi}(z',\om) &&= z'^{-3}\times z'^5J^2_{mR-1/2}(\om z')\, ,
\ee
$\rho\equiv 2/\sqrt{\lambda}$, $\zeta(z,z',\lambda,q^2,x_k)\equiv \log(\frac{zz'}{\sqrt{\lambda}}\frac{q^2}{x_k})$, and $g_{0}^2\equiv\frac{\kappa_{5}^2}{R^3}=4\pi^2/N_{c}^2$.

\section{Black hole with \newline\newline $e_R^2\tilde{\alpha}<\frac{1}{4}(mR)^2$ or $k_R^2<0$}

Near the black-hole the bulk fermions get modified in the infrared as illustrated in Fig.~\ref{fig_qbs}. The modification depends quantitatively
on their charge and mass near the horizon as we discussed earlier. For fermions with $e_R^2\tilde{\alpha}<\frac{1}{4}(mR)^2$ or $k_R^2<0$, the
modification is universal and follows from the reduced AdS$_2\times$ R$^3$ geometry. More specifically, for
hard R-probes with large $q^0$ in the DIS kinematics, only ${\cal G}_R(k^0, \vec k)$ is  modified close to the horizon,
since ${\cal G}_R(\omega_1, k+q)$ carries a large momentum and is mostly unmodified in the ultraviolet. In this regime,
the holographic fermions form a disc of radius $k_R$ in momentum space as we noted earlier, with large real
and imaginary parts. With this in mind, we have

\begin{widetext}
\be
\label{A7pure}
{\rm Im}\,\mathcal{G}_R^{11}(\om_1,k+q)\, {\rm Im}\,\mathcal{G}_{R}^{11}(k^0,\vk)\rightarrow
{\rm Tr}\,\Big((\sigma_{1}(k^0+q^0)-i\sigma_2(k_x+q_x)-\omega_1)\pi\,\delta((k+q)^2+\omega_1^2)
\times\,{\rm Im}\,\mathcal{G}^{11}_{R}(k^0,\vk)\Big)\nonumber\\
\ee
\end{widetext}
Here $\mathcal{G}^{11}_{R}=\mathcal{G}_R\, {\rm diag}(0, 1)$ is  given in (\ref{GR11}) for small $\omega$.
Again note the emerging non-Fermi liquid scaling for $\nu_k<\frac 12$ with the transition to a normal Fermi liquid
for $\omega\approx \omega_c$ as discussed earlier.
Using the vertex for pure AdS$_2$ (\ref{vertex-pure}), we can re-write (\ref{GRx3}) as

\begin{widetext}
\be
\label{GRxpure}
 {\rm Im}\,\tilde{G}^F_{xx}(q)= q^2C_{\theta}
(-1)\int_{0}^\infty\frac{d\om_1^2}{2} \int dk\,k^2C^2(\nu_{k})\,I_{z}^2(\om_1;q;k)\,{\rm Re}\,I_{k^0}(\om_1,q,x_{A})\,,\nonumber\\
\ee
\end{widetext}
with $C_\theta=1/(12\pi^2)$, and the real part of $I_{k^0}$ is

\begin{widetext}
\be
\label{I0pure}
{\rm Re}\, I_{k^0}(\om_1,q,x_{A})&&=
{\rm Re}\, \int_{-\infty}^{\infty}\frac{dk^{0}}{2\pi}\,{\rm Tr}\,\Big(\mathcal{G}_F^{11}(\om_1,k+q)\mathcal{G}^{11}_{F}(k^0,\vk)\Big)
\nonumber\\
&&={\rm Re}\,\int_{-|q^0|}^{0}\frac{dk^{0}}{2\pi}\,{\rm Tr}\,\Big((\sigma_{1}(k^0+q^0)-i\sigma_2(k_x+q_x)-\omega_1)\pi\,\delta((k+q)^2+\omega_1^2)\,{\rm Im}\,\mathcal{G}^{11}_{R}(k^0,\vk)\Big)
\nonumber\\
&&\approx \,\int_{-\omega_c}^{0}\frac{dk^{0}}{2\pi}(-1)\omega_1\pi\,\delta((k+q)^2+\omega_1^2)\,{\rm Im}\,C(\vec k)(k^0)^{2\nu_k}\,.
\ee
\end{widetext}
By first doing the integral over $\om_1$ in (\ref{GRxpure}) and using the delta function in (\ref{I0pure}), we finally obtain

\begin{widetext}
\be \label{GRxpure2}
{\rm Im}\,\tilde{G}^F_{xx}(q)
&&\approx \frac{C_{\theta}}4\int dk\,k^2\,C^2(\nu_{k})\,q^2I_{z}^2(\sqrt{s_k};q;k)\,\sqrt{s_k}\,\,{\rm Im}\,\left(\frac{-C(\vec k)(-\omega_c)^{2\nu_k+1}}{2\nu_k+1} \right)                                                                                                                                                                                                                                                                                                                                                                                                                                                                                                                                                                                                                                                                                                                                                                                                                                                                                                                                                                                                                                                                                                                                                                                                                                                                                                                                                                                                                                                                                                                                                                                                                                                                                                                                                                                                                                                                                                                                                                                                                                                                                                                                                                                                                                 \,,\nonumber\\
&&\approx \frac{C_{\theta}}4\Big(\frac{1}{q^2}\Big)^{\nu_{k}+\frac{1}{2}}\nonumber\\
&&\times\int dk\,k^2\,C^2(\nu_{k})\,C_{z}^2(\nu_{k})\,x_k^{\nu_{k}+5/2}(1-x_k)^{mR-1/2}\,_2F_1^2\Big(\frac{mR+\nu_{k}+2}{2},\frac{mR-\nu_{k}+1}{2},mR+\frac{1}{2},1-x_k\Big)\nonumber\\
&&\times\,\sqrt{s_k}\,\,{\rm Im}\,\left(\frac{-C(\vec k)(-\omega_c)^{2\nu_k+1}}{2\nu_k+1} \right)                                                                                                                                                                                                                                                                                                                                                                                                                                                                                                                                                                                                                                                                                                                                                                                                                                                                                                                                                                                                                                                                                                                                                                                                                                                                                                                                                                                                                                                                                                                                                                                                                                                                                                                                                                                                                                                                                                                                                                                                                                                                                                                                                                                                                                 \,,\nonumber\\
\ee
\end{widetext}
with $k^0$ fixed to $\omega_c$ in $s_k=-(k+q)^2$. We have defined
 $x_k={-q^2}/{2k\cdot q}$, $x_A={q^2}/{2E_A q_{x}}$, and made use of the DIS kinematics
 to approximate  $s_{k}\approx-q^2(1-{1}/{x_k})$, $|q^0|\approx q_x$.


 \vfil

\end{document}